%% file: paperdraft.tex
\newcommand*{\ATLASLATEXPATH}{./}
\documentclass[USenglish,cernpreprint,texlive=2011]{\ATLASLATEXPATH atlasdoc}

\usepackage{\ATLASLATEXPATH atlaspackage}
\usepackage{float}
\usepackage{comment}
\usepackage[utf8]{inputenc} 
\usepackage{\ATLASLATEXPATH atlasbiblatex}

\usepackage{\ATLASLATEXPATH atlascontribute}

\usepackage[BSM,process,xref]{\ATLASLATEXPATH atlasphysics}

\usepackage{pdflscape}
\usepackage{adjustbox}
\usepackage{paperdraft-defs}
\usepackage{a4wide}

\newcommand{\AtlasCoordFootnote}{
ATLAS uses a right-handed coordinate system with its origin at the nominal interaction point (IP)
in the center of the detector and the $z$-axis along the beam pipe.
The $x$-axis points from the IP to the center of the LHC ring,
and the $y$-axis points upwards.
Cylindrical coordinates $(r,\phi)$ are used in the transverse plane, 
$\phi$ being the azimuthal angle around the $z$-axis.
The pseudorapidity is defined in terms of the polar angle $\theta$ as $\eta = -\ln \tan(\theta/2)$.
Angular distance is measured in units of $\Delta R \equiv \sqrt{(\Delta\eta)^{2} + (\Delta\phi)^{2}}$.}

\AtlasTitle{Search for top squarks in final states with one isolated
lepton, jets, and missing transverse momentum in \sqrtS  $pp$
collisions with the ATLAS detector}
\date{\today}
\PreprintIdNumber{CERN-EP-2016-113}
\AtlasJournal{Phys.\ Rev.\ D.}
\AtlasAbstract{%
The results of a search for the stop, the supersymmetric partner of the top quark, in final states with one isolated electron or muon, jets, and missing transverse momentum are reported. The
search uses the 2015 LHC $pp$ collision data at a center-of-mass energy of \sqrtS recorded by the ATLAS detector and corresponding to an integrated luminosity of \ourLumi. 
The analysis targets two types of signal models:  gluino-mediated pair production of stops with a nearly mass-degenerate stop and neutralino; and direct pair production of stops, decaying to the top quark and the lightest neutralino. The experimental signature in both signal scenarios is similar to that of a top quark pair produced in association with large missing transverse momentum. No significant excess over the Standard Model background prediction is observed, and exclusion limits on gluino and stop masses are set at 95\% confidence level. The results extend the LHC Run-1 exclusion limit on the gluino mass up to 1460\,\GeV\ in the gluino-mediated scenario in the high gluino and low stop mass region, and add an excluded stop mass region from 745 to 780\,\GeV\ for the direct stop model with a massless lightest neutralino.
The results are also reinterpreted to set exclusion limits in a model of vector-like top quarks.}

\hypersetup{pdftitle={ATLAS document},pdfauthor={The ATLAS Collaboration}}

\begin{document}

\maketitle

\clearpage

\section{Introduction}
\label{sec:intro}

Supersymmetry (SUSY)~\cite{Golfand:1971iw,Volkov:1973ix,Wess:1974tw,Wess:1974jb,Ferrara:1974pu,Salam:1974ig} is a natural solution~\cite{Barbieri:1987fn,deCarlos:1993yy} to the hierarchy problem~\cite{Sakai:1981gr,Dimopoulos:1981yj,Ibanez:1981yh,Dimopoulos:1981zb}.
The top squark or stop ($\tilde{t}$), which is the superpartner of the top quark, is expected to be relatively light due to its large contribution to the Higgs boson mass radiative corrections~\cite{Inoue:1982pi,Ellis:1983ed}. 
For reasons such as gauge unification~\cite{Raby:2009sf} and the two-loop radiative corrections to the Higgs boson mass~\cite{Hardy:2013ywa,Arvanitaki:2013yja}, one may also expect 
a \TeV\ mass scale for the gluino (\gluino), the superpartner of the gluon.
A common theoretical strategy for avoiding strong constraints from the nonobservation of proton decay~\cite{Abe:2014mwa} is to introduce a multiplicative quantum number called $R$-parity.
If $R$-parity is conserved~\cite{Farrar:1978xj}, SUSY particles are produced in pairs and the lightest supersymmetric particle (LSP) is stable.
This analysis follows the typical assumption that the lightest neutralino\footnote{
The charginos \chinoOneTwopm\ and neutralinos \ninoOneTwoThreeFour\ are the mass eigenstates 
formed from the linear superposition of the charged and neutral SUSY partners of the Higgs and electroweak gauge bosons 
(higgsinos, winos and binos).} (\ninoone) is the LSP. Since the \ninoone\ interacts only weakly, it can serve as a candidate for dark matter~\cite{Goldberg:1983nd,Ellis:1983ew}.

\begin{figure}
\begin{center}
\includegraphics[width=0.4\textwidth]{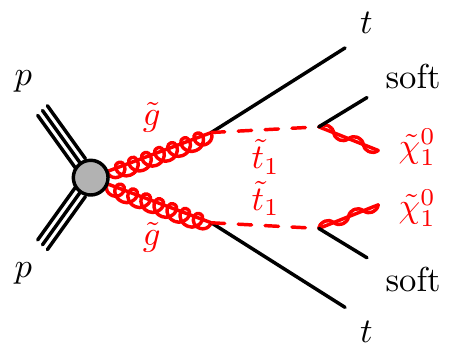}
\includegraphics[width=0.4\textwidth]{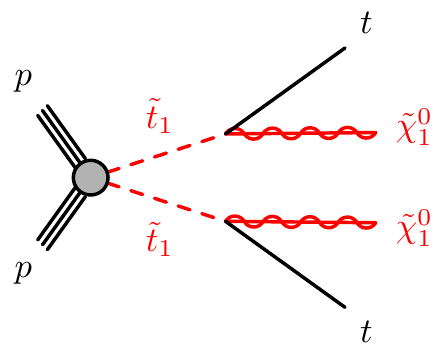}\hspace{10mm}
\caption{Diagrams illustrating the two considered signal scenarios.
Left: gluino-mediated stop pair production, 
where each stop decays to a low momentum (`soft') charm quark and the lightest neutralino.  
Right: stop pair production, where each stop decays to the top quark and the lightest neutralino (\ninoone).
For simplicity, no distinction is made between particles and antiparticles.
 }
\label{fig:diagram}
\end{center}
\end{figure}

This paper presents a search targeting the lighter stop\footnote{The superpartners of the left- and right-handed top quarks, \stopL\ and \stopR, mix to form the two mass eigenstates \stopone\ and \stoptwo, where \stopone\ is the lighter one.} (\stopone) in two scenarios:  
gluino-mediated pair production of the \stopone\ with a small \stopone-LSP mass splitting, 
and direct pair production of the \stopone, both illustrated by the diagrams in Fig.~\ref{fig:diagram}.
The former scenario refers to pair production of gluinos, each decaying to the top quark and the \stopone.  In this scenario, the mass difference between the gluino and the \stopone\ is assumed to be well above the top quark mass, while the mass difference between the \stopone and the LSP is assumed to be significantly smaller than the $W$ boson mass.  As a result, the visible \stopone\ decay products have low momentum, typically below the reconstruction
and identification thresholds. This scenario is motivated by the dark matter relic density, which is generally too large in the Minimal Supersymmetric Standard Model~\cite{Fayet:1976et,Fayet:1977yc} but can be regulated by coannihilation of the stop and the neutralino~\cite{deSimone:2014pda}.  
In the second scenario, the two directly produced \stopone\ are each assumed to decay to the top quark and the LSP. This model is interesting as it is independent of the gluino mass, which is more weakly constrained by naturalness arguments than the stop mass.

Experimentally, the final states of the two scenarios are similar~\cite{Nachman:2015aaa},
 and the detector signature consists of the decay products of a pair of top quarks\footnote{Due to the Majorana nature of the gluino, in the gluino-mediated model, each of the two `visible' top quarks can independently be a top or an antitop quark.
Hereafter, the term \ttbar\ can be taken to refer to any combination of $t$ and $\bar{t}$.}
and large missing transverse momentum (\Ptmiss, where the magnitude is referred to as \met) from the two LSPs: $\ttbar+\met$.
The main difference between the two scenarios is that the production cross-section for gluino pairs is about a factor 50 higher than for \stopone pairs of the same mass due to the additional spin and color states.
The results are also reinterpreted in a model of strong-interaction direct pair production of vector-like top quarks $T$ (referred to as VLQ)~\cite{Schmaltz:2005ky,Agashe:2004rs,Aguilar:1553376}, for which the decay mode $T\rightarrow tZ$ with $Z \to \nu \bar{\nu}$  has a signature similar to that of direct stop pair production with $\stopone \to t \ninoone$.

The analysis presented here -- which 
is based on previous ATLAS searches for the same signature~\cite{SUSY-2013-15,SUSY-2012-06} --
targets the one-lepton final state 
where the $W$ boson from one of the top quarks decays to an electron or muon (either directly or via a $\tau$ lepton) and the $W$ boson from the other top quark decays hadronically. 
The dominant Standard Model (SM) background processes are: the production of \ttbar\ ; the associated production of a top quark and a $W$ boson (single top $Wt$);
 $\ttbar + Z(\to \nu\bar{\nu}) $;
 and the associated production of $W$ bosons and jets ($W$+jets).
The search uses the ATLAS data collected in proton-proton ($pp$) collisions in 2015 corresponding to an integrated luminosity of \ourLumi\ at a center-of-mass energy of $\sqrt{s}=13$~\TeV.  
The ATLAS Run-1 searches for gluino-mediated stop production and direct stop pair production are summarized in Refs.~\cite{SUSY-2014-06} and~\cite{SUSY-2014-07}, respectively.  Run-1 searches for VLQ production can be found in Refs.~\cite{EXOT-2013-16,EXOT-2013-17,EXOT-2013-18}. The CMS Collaboration has performed similar searches for gluino-mediated stop production~\cite{Khachatryan:2016zcu,}, direct stop pair production~\cite{
CMS-SUS-14-001,Khachatryan:2016pup,CMS-SUS-13-011,CMS-SUS-13-004,CMS-SUS-13-019,Khachatryan:2016xvy}, and VLQ production~\cite{Khachatryan:2015oba}.

This document is organized as follows.  The ATLAS detector, dataset, and trigger are described in Section~\ref{sec:detector}, and the corresponding set of simulations are detailed in Section~\ref{sec:simulation}.  Section~\ref{sec:eventselection} presents 
the reconstruction and selection of physics objects and the construction of discriminating variables.  
These variables are used in Section~\ref{sec:SRs} to construct the signal  event selections.  The background estimation procedure (Section~\ref{sec:background}) and systematic uncertainties (Section~\ref{sec:systematics}) are described before the results are presented in Section~\ref{sec:result}.  Section~\ref{sec:conclusion} contains concluding remarks.

\section{ATLAS Detector and Dataset}
\label{sec:detector}


The ATLAS detector~\cite{PERF-2007-01} is a multipurpose particle physics detector 
with nearly $4\pi$ coverage in solid angle around the collision point.\footnote{\AtlasCoordFootnote}
It consists of an inner tracking detector (ID),  surrounded by a superconducting solenoid providing a \SI{2}{\tesla} axial magnetic field, a system of calorimeters, and a muon spectrometer (MS) incorporating three large superconducting toroid magnets.  
The ID provides charged-particle tracking in the range $|\eta| < 2.5$ using three technologies: silicon pixel and silicon microstrip tracking detectors, and a transition radiation tracker. During the LHC shutdown between Run 1 and Run 2, a new innermost layer of silicon pixels was added, which improves 
the track impact parameter resolution and vertex position resolution~\cite{ATL-PHYS-PUB-2015-051}.  
High-granularity electromagnetic and hadronic calorimeters cover the region $|\eta| < 4.9$. 
The central hadronic calorimeter is a sampling calorimeter with scintillator tiles as the active medium and steel absorbers. All the electromagnetic calorimeters, as well as the endcap and forward hadronic calorimeters, are sampling calorimeters with liquid argon as the active medium and lead, copper, or tungsten absorber.
The MS consists of three layers of high-precision tracking chambers with coverage up to $|\eta|=2.7$ and dedicated chambers for triggering in the region $|\eta|<2.4$. 
Events are selected by a two-level trigger system: the first level is a hardware-based system and the second is a software-based system.

The 2015 LHC collision data used in this analysis has a mean number of additional $pp$ interactions per bunch crossing (pileup) of approximately 14, and 
a bunch spacing of $25$\,ns.
Following requirements based on beam and detector conditions and data quality, the  dataset corresponds to an integrated luminosity of 3.2\,\ifb\ with an associated uncertainty of 5\%. The uncertainty is derived following the same methodology as that detailed in Ref.~\cite{DAPR-2011-01}.
Events used for this search were recorded using a trigger logic that accepts events with \met, calibrated to the electromagnetic scale, above $70$\,\GeV.   The trigger is more than 95\% efficient for events passing an offline-computed $\met > 200$\,\GeV\ requirement and is $>99$\% efficient for events passing all signal selections.  An additional data sample used to estimate one of the background processes was recorded with a trigger requiring a photon with transverse momentum $\pt > 120$\,\GeV, which is $>99\%$ efficient for the offline photon selection described in Section~\ref{sec:eventselection}.

\section{Monte Carlo Simulations }
\label{sec:simulation}

Samples of Monte Carlo (MC) simulated events are used for the description 
of the background and to model the SUSY signals.  Several matrix 
element (ME) generators are combined with parton shower (PS) and hadronization generators.  
Signal SUSY samples are generated at leading order (LO) with 
\MGaMC~2~\cite{Alwall:2014hca} while VLQ signal samples are generated at LO with \PROTOS v2.2~\cite{AguilarSaavedra:2009es,Aguilar:soft}.
All signal samples are interfaced with \PYTHIA 
8.186~\cite{Sjostrand:2007gs}.
Background samples use one of three setups:

\begin{itemize}
\item \MGaMC~v2~interfaced with \PYTHIA 8 or \HERWIGpp using the CKKW-L~\cite{Lonnblad:2011xx} or the MC@NLO method for matching a LO or next-to-leading-order (NLO) ME to the PS, respectively.
\item \powbox~\cite{Alioli:2010xd,Re:2010bp,Frixione:2007nw,Frederix:2012dh,Alioli:2009je} interfaced to \PYTHIA 6~\cite{Sjostrand:2006za} or \HERWIGpp using the \POWHEG method~\cite{Nason:2004rx,Frixione:2007vw} for matching the NLO ME to the PS.
\item \SHERPA 2.1.1~\cite{Gleisberg:2008ta} using Comix~\cite{Gleisberg:2008fv} and OpenLoops~\cite{Cascioli:2011va} ME generators interfaced with the \SHERPA parton shower~\cite{Schumann:2007mg}.
\end{itemize}

The CT10~\cite{Lai:2010vv} NLO parton distribution function (PDF) set is used for ME calculations with \SHERPA and \powbox\ and the NNPDF2.3~\cite{Ball:2012cx} PDF set is used for samples generated with \MGaMC, except for the NLO samples, which use either CT10 or NNPDF3.0~\cite{Ball:2014uwa}.  The {\sc CTEQ6L1}~\cite{Pumplin:2002vw} LO PDF set along with the P2012~\cite{Skands:2010ak} set of underlying-event tuned parameters (UE tune) is used for \PYTHIA 6; the NNPDF2.3 LO PDF set and the A14 UE tune~\cite{ATL-PHYS-PUB-2014-021} is used for \PYTHIA 8; and the CT10 PDF set with the default UE tune provided by the authors of \SHERPA is used for the \SHERPA samples.  The samples produced with \MGaMC, \powbox, and \PROTOS all use {\sc EvtGen} v1.2.0~\cite{EvtGen} for the modeling of $b$-hadron decays.  The simulation setup is summarized in Table~\ref{tab:mc_samples1} and more details can be found in Refs.~\cite{ATL-PHYS-PUB-2016-004,ATL-PHYS-PUB-2016-003,ATL-PHYS-PUB-2016-005,ATL-PHYS-PUB-2016-002} for $\ttbar$ and single top, $W/Z$+jets, dibosons, and $\ttbar+W/Z$, respectively.   Additional samples aside from those shown in Table~\ref{tab:mc_samples1} are used to assess theoretical modeling uncertainties and are discussed in Section~\ref{sec:systematics}.

\begin{table}[h!]
\centering
\begin{tabular}{| l | lllll |}
\hline
Process & ME generator & ME  & PS and & UE  & Cross-section\\
 &  &  PDF & Hadronization & tune & order\\
\hline
$\ttbar$ & \powbox~v2 & CT10& \PYTHIA 6 & P2012 &NNLO+NNLL~\cite{Czakon:2013goa,Czakon:2012pz,Czakon:2012zr,Baernreuther:2012ws,Cacciari:2011hy,Czakon:2011xx} \\ 
Single top& \powbox & CT10 & \PYTHIA 6 & P2012 &NNLO+NNLL~\cite{Kidonakis:2011wy,Kidonakis:2010ux,Kidonakis:2010tc} \\ 
$W/Z$+jets & \SHERPA 2.1.1 & CT10 & \SHERPA & Default & NNLO~\cite{Catani:2009sm}\\ 
Diboson & \SHERPA 2.1.1 & CT10 & \SHERPA & Default & NLO\\ 
$\ttbar+W/Z$ & \MGaMC~2.2.2 & NNPDF2.3 & \PYTHIA 8 & A14 & NLO~\cite{Alwall:2014hca}\\ 
$\ttbar+\gamma$& \MGaMC~2.2.3 & CTEQ6L1 & \PYTHIA 8 & A14 & NLO~\cite{Alwall:2014hca}\\ 
SUSY signal & \MGaMC~2.2.2 & NNPDF2.3 & \PYTHIA 8 & A14 & NLO+NLL~\cite{Borschensky:2014cia}\\ 
VLQ signal & \PROTOS v2.2 & NNPDF2.3 & \PYTHIA 8  &  A14  & NNLO+NNLL~\cite{Czakon:2013goa,Czakon:2012pz,Czakon:2012zr,Baernreuther:2012ws,Cacciari:2011hy,Czakon:2011xx}  \\
\hline
\end{tabular}
\caption{Overview of the nominal simulated samples. }
\label{tab:mc_samples1}
\end{table}

In the gluino-mediated production the stop is assumed to decay via  $\stopone 
\rightarrow c+\ninoone$ with a 100\% branching ratio and with a default mass 
splitting $m_{\stopone} - m_{\ninoone}=5$\,\GeV. Alternative samples with 
larger mass splitting and/or replacing the two-body stop decay by a four-body 
stop decay $\stopone \rightarrow bff'\ninoone$, where $ff'$ is a fermion-antifermion pair, are produced for additional studies.
The gluinos and stops are assumed to decay promptly.
In the direct stop pair production samples, the $\stopone$ is chosen to be mostly the partner of the right-handed top quark\footnote{
The \stopR\ component is given by the the off-diagonal entry of the stop mixing matrix. 
The \stopone\ decays in the direct stop pair production samples are performed by \PYTHIA\ and produce unpolarized top quarks. 
The events are reweighted to obtain a stop mixing equivalent to 
a matrix with on-diagonal entries of approximately $0.55$ and off-diagonal entries of approximately $\pm 0.83$.
The event weights depend on the angular distributions of the top decay products~\cite{Low:2013aza}.
}  and the $\ninoone$ to be a pure bino. 
This choice is consistent with a large branching ratio for the given \stopone\ decay.
Different hypotheses for the left/right mixing in the stop sector and the nature of the neutralino lead to different acceptance values. The acceptance is affected because the polarization of the top quark changes as a function of the field content of the supersymmetric particles, which impacts the boost of the lepton in the top quark decay.
Signal grids are generated for both the gluino and direct stop pair production models. 
The spacing between grid points in the gluino-stop and stop-neutralino mass planes vary
between 25 and 100\,\GeV.

All the MC samples are normalized to the highest-order (in \alphas) cross-section available, as indicated in the last column of Table~\ref{tab:mc_samples1}.  The cross-sections for the pair and single production of top quarks as well as for the signal processes also include resummation of soft gluon emission to next-to-next-to-leading-logarithmic (NNLL) and next-to-leading-logarithmic (NLL) accuracy, respectively.  
As is described in Section~\ref{sec:back:CRs:TZCR}, it is important 
that the simulated $\ttbar+\gamma$ and $\ttbar+Z$ events are as similar as possible.
Therefore, a small 4\% correction is applied to the $\ttbar+\gamma$ cross-section to account for a different PDF set, factorization/renormalization scale, and number of partons from the matrix element.\footnote{
The $\ttbar+\gamma$ sample uses a fixed factorization/renormalization scale of $2\times m_\text{top}$ with no extra partons in the ME. The $\ttbar+Z$ sample uses the default $\sum \mt$ scale
and is generated with up to two partons. 
The top decay is performed in \MGaMC~for $\ttbar+\gamma$ to account for hard photon radiation from the top decay products, which is a $\sim15$\% effect for $\pt^\gamma\sim120$ \GeV~\cite{Melnikov:2011ta}.
} The same NLO QCD $K$-factor is then applied to the $\ttbar+\gamma$ process as is used for the $\ttbar+Z(\rightarrow\nu\bar{\nu})$ process~\cite{Alwall:2014hca}.  This choice is motivated by the similarity of QCD calculations for the two processes as well as empirical studies of the ratio of $K$-factors computed as a function of the boson \pt.  Further information about the $K$-factor and its uncertainty is given in Section~\ref{sec:systematics}.
The cross-sections for the \ttbar, $W$+jets, and $Wt$ processes are used for cross-checks and optimization studies, while for the final results these processes are normalized to data in control regions.

All background samples, except for the $\ttbar+\gamma$ sample, are processed with the full simulation of the ATLAS detector~\cite{SOFT-2010-01} based on \GEANT 4~\cite{Agostinelli:2002hh}. 
The signal samples and the $\ttbar+\gamma$ sample are processed with a fast simulation~\cite{ATL-PHYS-PUB-2010-013} of the ATLAS detector with parameterized showers in the calorimeters.  
All samples are produced with varying numbers of simulated minimum-bias interactions generated with \PYTHIA 8 overlaid on the hard-scattering event to account for pileup from multiple $pp$ interactions in the same or nearby bunch crossings. The average number of interactions per bunch crossing is reweighted to match the distribution in data.  %
Furthermore, the simulated samples are reweighted to account for small differences in the efficiencies of physics-object reconstruction and identification with respect to those measured in data. %

\section{Event Reconstruction and Selection}
\label{sec:eventselection}

All events must satisfy a series of quality criteria before being considered for further use. The reconstructed primary vertex with the highest $\sum_\text{tracks} \pt^2$ must have at least two associated tracks.
In this analysis, physics objects are labeled as either {\it baseline} or {\it signal} depending on various quality and kinematic requirements, where the latter label describes a tighter selection of the former. Baseline objects are used to distinguish between the physics objects in the event and to compute the missing transverse momentum. Baseline leptons (electrons and muons) are also used to apply a second-lepton veto
to suppress dilepton \ttbar\ and $Wt$ events.

Electron candidates are reconstructed from electromagnetic calorimeter cell clusters that are matched to ID tracks. Baseline electrons are required to have $\pT > 7$\,\GeV, $|\eta| < 2.47$, and satisfy `VeryLoose' likelihood identification criteria that are
defined following the methodology described in Ref.~\cite{ATL-PHYS-PUB-2015-041}. Signal electrons must pass all baseline requirements and in addition have $\pT > 25$\,\GeV, satisfy the `Loose' likelihood identification criteria in Ref.~\cite{ATL-PHYS-PUB-2015-041}, and have impact parameters with respect to the reconstructed primary vertex along the beam direction ($z_0$) and in the transverse plane ($d_0$) that satisfy $| z_0 \sin \theta| < 0.5$\,mm and $|d_0|/\sigma_{d_0} < 5$, where $\sigma_{d_0}$ is the uncertainty of  $d_0$. Furthermore, signal electrons must be isolated, where the criteria use track-based information to obtain a 99\% efficiency that is independent of $p_\text{T}$, as derived from $Z\rightarrow \ell\ell$ MC samples and confirmed in data.

Muons are reconstructed from combined tracks that are formed from ID and MS tracks, ID tracks matched to MS track segments, standalone MS tracks, or ID tracks matched to an energy deposit in the calorimeter compatible with a minimum-ionizing particle (referred to as calo-tagged muon)~\cite{Aad:2016jkr}. Baseline muons are required to have $\pt>6$\,\GeV, $|\eta|<2.7$, and satisfy the `Loose' identification criteria described in Ref.~\cite{Aad:2016jkr}. Signal muons must pass all baseline requirements and in addition have $\pt > 25$\,\GeV, and have impact parameters $| z_0 \sin \theta| < 0.5$\,mm and $|d_0|/\sigma_{d_0} < 3$. Furthermore, signal muons must be isolated according to isolation criteria similar to those used for signal electrons, yielding the same efficiency. %

Photon identification is not used in the main event selection, and photons give rise to extra jet or electron candidates.
Photons must be identified, however, for the $\ttbar + \gamma$ sample that is used in the data-driven estimation of the $\ttbar + Z$ background.
In this case, photon candidates are reconstructed from calorimeter cell clusters and are required to satisfy the `Tight' identification criteria described in Ref.~\cite{ATL-PHYS-PUB-2011-007}.  Furthermore, photons are required to have $\pt>125$\,\GeV\ and $|\eta|<2.37$, excluding 
the barrel-endcap calorimeter transition in the range $1.37 < |\eta| < 1.52$, so that the photon trigger is fully efficient.  
Photons must further satisfy isolation criteria based on both track and calorimeter information.

Jet candidates are built from topological clusters~\cite{ATL-LARG-PUB-2008-002,PERF-2011-03} in the calorimeters using the anti-$k_t$ algorithm with a jet radius parameter $R=0.4$~\cite{Cacciari:2008gp}. Jets are corrected for contamination from pileup using the jet area method~\cite{Cacciari:2007fd,Cacciari:2008gn,PERF-2014-03} and then calibrated to account for the detector response~\cite{PERF-2012-01,ATL-PHYS-PUB-2015-015}.  
Jets in data are further calibrated based on \textit{in situ} measurements of the jet energy scale.
Baseline jets are required to have $\pt > 20$\,\GeV. Signal jets must have $\pt > 25$\,\GeV\ and $|\eta|<2.5$.  
Furthermore, signal jets with $\pT < 50$\,\GeV\ are required to satisfy criteria, implemented in the jet vertex tagger algorithm~\cite{PERF-2014-03}, designed to reject jets originating from pileup. Events containing a jet that does not pass specific jet quality requirements are vetoed from the analysis in order to suppress detector noise and noncollision backgrounds~\cite{DAPR-2012-01,ATLAS-CONF-2015-029}.
Jets resulting from $b$-quarks (called $b$-jets) are identified using the MV2c20 $b$-tagging algorithm, which is based on quantities such as impact parameters of associated tracks and reconstructed secondary vertices~\cite{ATL-PHYS-PUB-2015-022,ATL-PHYS-PUB-2015-039,PERF-2012-04}.
This algorithm is used at a working point that provides 77\% $b$-tagging efficiency in simulated $\ttbar$ events.
The choice of working point was optimized for this analysis and corresponds to a rejection factor of about 140 for light-quark flavors and gluons and about 5 for charm jets.  
Jets and associated tracks are also used to identify hadronically decaying $\tau$ leptons using the `Loose' identification criteria described in Refs.~\cite{ATL-PHYS-PUB-2015-025,ATL-PHYS-PUB-2015-045}, which have a 60\% and 50\% efficiency for reconstructing $\tau$ leptons decaying into one and three charged pions, respectively.  These $\tau$ candidates are required to have one or three associated tracks, with total electric charge opposite to that of the selected electron or muon, 
$\pt>20$ \GeV, and $|\eta|<2.5$.  This $\tau$ candidate \pT\ requirement is applied after a dedicated energy calibration~\cite{ATL-PHYS-PUB-2015-045}.

The missing transverse momentum is reconstructed from the negative vector sum of the transverse momenta of baseline electrons, muons, jets, and a {\it soft-term} built from high-quality tracks that are associated with the primary vertex but not with the baseline physics objects~\cite{ATL-PHYS-PUB-2015-027,ATL-PHYS-PUB-2015-023}.  For the event selections requiring photons, the calibrated photon is directly included in the \met calculation.  In all other cases, photons and hadronically decaying $\tau$ leptons are not explicitly included but enter as jets or electrons, or via the soft-term.  

To avoid labeling the same detector signature as more than one object, an overlap removal procedure is applied. The procedure is tailored for this analysis and optimized using simulation.
Table~\ref{tab:OR} summarizes the procedure. 
Given a set of baseline objects, the procedure checks for overlap based on a minimal distance $\DeltaR$ between pairs of objects. 
For example, if a baseline electron and a baseline jet are found with $\Delta R < 0.2$, then the electron is retained (as stated in the `Precedence' row) and the jet is discarded, unless the jet is $b$-tagged (as stated in the `Condition' row) in which case the electron is assumed to stem from a heavy-flavor decay and is hence discarded while the jet is retained.
If the `$\Delta R$<' requirement in Table~\ref{tab:OR} is not met, then both objects under consideration are kept.
The order of steps in the procedure is given by the columns in Table~\ref{tab:OR}, which are executed from left to right. 
The second ($e j$) and the third ($\mu j$) steps of the procedure ensure that leptons and jets have a minimum $\Delta R$ separation of $0.2$. Therefore, 
the fourth step ($\ell j$) only has an effect for $\Delta R > 0.2$.
The steps involving a photon are not applied in the main event selection, but only for the event selection where photons are identified. 
For the remainder of the paper, all baseline and signal objects are those that have survived the overlap removal procedure.

\begin{table}
{\small
  \centering
  \vspace{3mm}
{\renewcommand{\arraystretch}{1.3}%
  \begin{tabular}{| c | ccccccc |}
	\hline
    Object 1 & $e$     & $e$ & $\mu$ & $\ell$  & $\gamma$& $\gamma$ & $\tau$        \\
    Object 2 & $\mu$ & $j$ & $j$ & $j$ & $j$ & $e$ & $e$\\
	\hline
    $\Delta R<$ & 0.01 & 0.2 & 0.2 &$\min\left(0.4,0.04+\frac{10}{\pt^\ell/\text{GeV}}\right)$ & 0.2 & 0.1 & 0.1 \\
    	\hline
    Condition & calo-tagged $\mu$ & $j$ not $b$-tagged & $j$ not $b$-tagged and & -- & -- & -- & --\\
    & & & $\left(n_\text{track}^j<3 \ \text{or} \ \frac{\pt^\mu}{\pt^j }> 0.7\right)$ &&&&\\
	\hline
    Precedence & $e$ & $e$ & $\mu$ & $j$ & $\gamma$ & $e$ & $e$ \\
\hline
  \end{tabular}
}
  \caption{Overlap removal procedure.  The first two rows list the types of overlapping objects: electrons ($e$), muons ($\mu$), electron or muon ($\ell$), jets ($j$), photons ($\gamma$), and hadronically decaying $\tau$ lepton ($\tau$). All objects refer to the baseline definitions, except for $\gamma$ and $\tau$ where no distinction between baseline and signal definition is made.  The third row specifies when an object pair is considered as overlapping, the fourth row describes an optional condition, and the last row lists which label is given to the ambiguous object.  %
More information is given in the text.  %
  }
  \label{tab:OR}
  }
\end{table}

Large-radius jets are clustered from all signal (small-radius $R=0.4$) jets using the anti-$k_t$ algorithm with $R=1.0$ or $1.2$.
To reduce the impact of soft radiation and pileup, the large-radius jets are groomed using reclustered jet trimming,
where constituents with \pt less than 5\% of the ungroomed jet \pt are removed~\cite{SUSY-2013-04,SUSY-2013-16,Nachman:2014kla,Krohn:2009th}.
Electrons and muons are not included in the reclustering, since it was found that including them increases the background acceptance more than the signal efficiency. 
Large-radius jets are not used in the overlap removal procedure; however, the signal jets that enter the reclustering have passed the overlap removal procedure described above.
The analysis uses a large-radius jet mass, where the squared mass is defined as the square of the four-vector sum of the constituent (small-radius) jets' momenta.

All events are required to have $\met > 200$ \GeV, exactly one signal lepton, and no additional baseline leptons,
as well as at least four signal jets.  In addition, the events must have 
a transverse mass\footnote{The transverse mass $\mt$ is defined as $\mt^2=2p_{\text{T}}^{\text{lep}}\met[1-\cos(\Delta\phi)]$, where $\Delta\phi$ is the azimuthal angle between the lepton and the missing transverse momentum direction.  The quantity $p_\text{T}^\text{lep}$ is the transverse momentum of the charged lepton.} of the signal lepton and the missing transverse momentum
satisfying $\mt >30$\,\GeV, and have an azimuthal angle between leading or subleading jet and the missing transverse momentum of $\minDeltaPhi >0.4$ with $i\in\{1,2\}$.
The events must further pass an $\HTmissSig\ > 5$ requirement, where $\HTmissSig =(\HTmiss - 100 \GeV)/\sigma_{\HTmiss}$.  The variable $\HTmiss$ is the magnitude of
the negative vector sum of the transverse momenta of signal jets and the signal lepton; the resolution $\sigma_{\HTmiss}$ is computed using the per-event jet energy resolution uncertainties (more details are given in Refs.~\cite{SUSY-2013-15,Nachman:2013bia}).
The latter three event selection criteria for \mt, \minDeltaPhi, and \HTmissSig suppress multijet processes with misidentified or nonprompt leptons and mismeasured \met\ to a negligible level.
With the above event selection, the dominant backgrounds are \ttbar events with at least one leptonically decaying $W$ boson, and $W$+jets production.  
A powerful technique for suppressing these background processes is to require \mt\ to be greater than the $W$ boson mass.
For example, an $\mt > 120$\,\GeV\ requirement removes more than $90$\% of \ttbar\ and $W$+jets events that pass the above event selection.

One of the dominant contributions to the residual background is from $\ttbar$ production where both $W$ bosons decay leptonically, or one $W$ boson decays leptonically and the other via a hadronic $\tau$ decay.
A series of additional variables, described in detail in Ref.~\cite{SUSY-2013-15}, 
are used to discriminate between this background and the 
signal processes.
The \mTopChi\ variable is the invariant mass of the three jets in the event most compatible 
with the hadronic decay products of a top quark, 
where the three jets are selected by a $\chi^2$-minimization 
including the jet momenta and energy resolutions.
The asymmetric \mtTwo (\amtTwo)~\cite{Barr:2009jv,Konar:2009qr,Bai:2012gs,Lester:2014yga} and \mtTwoTau are both variants of the variable 
\mtTwo~\cite{Lester:1999tx}, a generalization of the transverse mass applied to signatures where
two particles are not directly detected. The \amtTwo\ variable targets 
dileptonic \ttbar\ events where one lepton is not 
reconstructed, while the \mtTwoTau variable targets \ttbar\ events where
one of the two $W$ bosons decays via a hadronically decaying $\tau$ lepton.
The \topness~\cite{Graesser:2012qy} variable is based on minimizing a $\chi^2$-type function quantifying the
compatibility with a dileptonic \ttbar\ event where one lepton is not reconstructed.
Furthermore, the mass of large-radius jets is useful when the boost of the 
top quark is significant.

An important change from the Run-1 suite of tools is the treatment of hadronically decaying $\tau$ candidates in the \mtTwoTau variable.  
Events are removed if one of the selected jets is additionally identified as a hadronic $\tau$ candidate, with a corresponding $\mtTwoTau < 80$\,\GeV, where
\mtTwoTau uses the signal lepton and hadronic $\tau$ candidate as the two visible objects.
For an event selection with a $\met > 200$\,\GeV\ requirement, this hadronic $\tau$ veto removes approximately 40\% of simulated \ttbar\ events where 
one $W$ boson decays leptonically and the other decays via a hadronically decaying $\tau$ lepton. 
For the considered signal models, the veto removes 1\% of the events. %
The $\tau$ veto is applied in all following event selections except those defining the $\ttbar+Z$ control region (since the veto would remove only about 1\% of the events in this region).
%

\section{Signal Regions}
\label{sec:SRs}

%
Three signal event selections (called signal regions, or SR1--3) are constructed using the set of discriminating variables 
described in Section~\ref{sec:eventselection}. %
The three signal regions are optimized, before looking at the data, to maximize the discovery sensitivity using three benchmark signal models from the gluino-mediated stop models, each representing a distinct phenomenology. The benchmark models are defined by (\gluino, \ninoone) masses of (1100, 800), (1250, 750), and (1400, 400)\,\GeV\ for SR1, SR2, and SR3, respectively. 
The benchmark model for SR1 has a production cross-section and kinematic properties similar to those of a direct stop model with (\stopone, \ninoone) masses of about (600, 260)\,\GeV, while
the benchmark models for SR2 and SR3 cannot be directly mapped to have both the same cross-sections and similar kinematic properties. 
As a consequence, SR2 and SR3 have reduced sensitivity to direct stop models. 

The three signal regions are characterized by increasing \met\ requirements.  The SR1 benchmark has the softest \met spectrum and the momentum of the hadronically decaying top quark is typically not sufficient to capture all of the decay products inside a single large-radius jet.  As a result, the top quark mass computed using the \mTopChi\ variable which is based on small-radius jets is useful for rejecting dileptonic $\ttbar$ and other background events without a top quark that has hadronic decay products.  In contrast, the  boost of the hadronically decaying top quarks in the SR2 and SR3 benchmarks is often sufficient to capture all decay products inside a single large-radius jet.  The angular separation between the decay products scales with the inverse of the momentum.  Therefore, the optimal large-radius jet cone size is found to be larger for SR2 ($R=1.2$) than for SR3 ($R=1.0$).  Additional requirements on \topness~and \amtTwo further reduce the dileptonic $\ttbar$ background.  Background events without a high-\pt top quark that decays leptonically are suppressed by using a requirement on the $\Delta R$ between the highest-\pt $b$-tagged jet and the signal lepton.  The signal regions have additional requirements on the \mt and \HTmissSig\ variables to further exploit the large genuine \met\ from undetected neutralinos.  
A requirement of at least one $b$-tagged jet is used in each of SR1--3 in order to reduce the $W$+jets and diboson backgrounds.

The signal region definitions are summarized in Table~\ref{tab:back:SR:CRs}.
The signal regions are not mutually exclusive.
%

\section{Background Estimates}
\label{sec:background}

The dominant background processes are \ttbar, single 
top $Wt$, $\ttbar+Z(\rightarrow\nu\bar{\nu})$, and $W$+jets. 
Most of the \ttbar and $Wt$ events in the signal regions have both $W$ bosons decay leptonically 
(one of which is `lost', meaning it is either not reconstructed, not identified, or removed by the overlap removal procedure)
or one $W$ boson decays leptonically and the other via a hadronically decaying $\tau$ lepton.
Other background processes considered are semileptonic \ttbar, dibosons (denoted by VV in figure legends), $\ttbar+W$, $Z$+jets, and multijet events.
The \ttbar background is shown separately in the three decay components discussed above, which are referred to as 2L, 1L1$\tau$, and 1L respectively.\footnote{The letter L is used to denote an electron or muon, including those from a leptonic $\tau$ decay; the $\tau$ symbol is used to denote a hadronic $\tau$ decay.}  The combined \ttbar+$W$ and \ttbar+$Z$ background is referred to as \ttbar+$V$.

The main background processes are estimated by isolating each of them in a 
dedicated control region (CR), described in Section~\ref{sec:back:CRs}, normalizing simulation to match data in a 
simultaneous fit.  The fit is performed separately for each SR with the associated CRs. The background modeling as predicted by the fits is tested in a series of validation regions (VR), discussed in Section~\ref{sec:back:VRs}.  Figure~\ref{fig:back:SRCRVR} 
schematically illustrates the setup for one example SR and its associated CRs and VRs.  The CRs for $Wt$ and $\ttbar +Z$ are new with respect to the Run-1 analysis. %

\begin{figure}
\begin{center}
\includegraphics[width=0.75\textwidth]{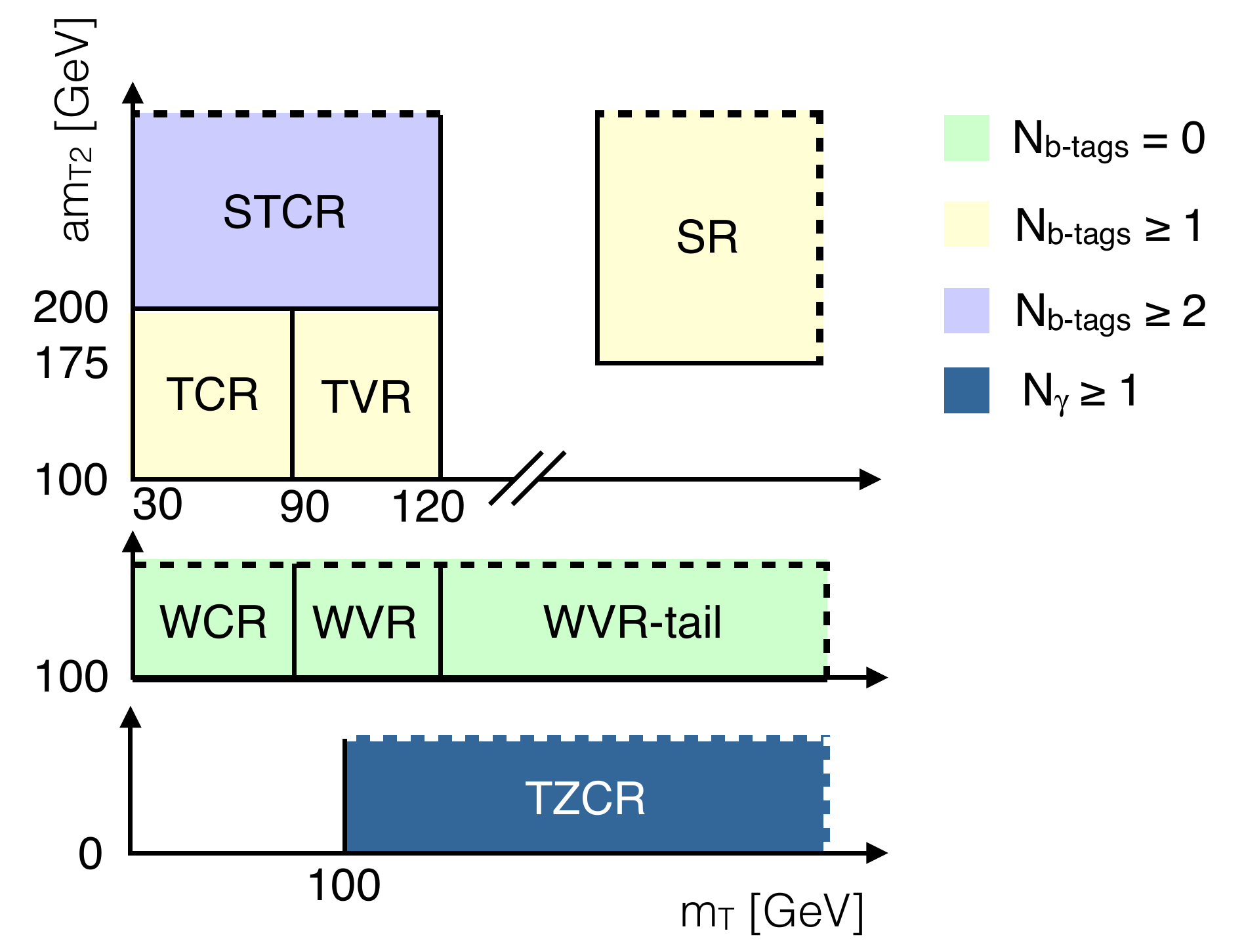}
\end{center}
\caption{A schematic diagram for the various event selections used to estimate 
and validate the background model and then search for stop production. Solid lines 
indicate kinematic boundaries while dashed lines indicate that the events can extend beyond the boundary.  CR, VR, and SR stand for control region, validation region, and signal region, respectively.  T, ST, TZ, and W stand for $\ttbar$, single top, $\ttbar+Z$, and $W$+jets, respectively.  
\label{fig:back:SRCRVR}}
\end{figure}

The multijet background is estimated from data using a fake-factor method~\cite{HIGG-2013-13}. The contribution is found to be negligible. All other (small) backgrounds are determined from simulation, normalized to the most accurate theoretical cross-sections available. The $Z$+jets background is found to be negligible.

\label{sec:back:fit}

\subsection{Control Regions}
\label{sec:back:CRs}
A series of control regions are defined as event selections that are kinematically close to the signal regions but with a few key variable requirements inverted to significantly reduce signal contamination and enhance the yield and purity of a particular background.  These control regions are then used to constrain the background normalization.  Each of the three signal regions has a dedicated control region for each of the following background processes: \ttbar\ (TCR), $W$+jets (WCR), single top (STCR), and $\ttbar+W/Z$ (TZCR). The general strategy in constructing the control regions is to {\it invert} the transverse mass requirement from a high threshold to a low window. The requirements on several variables are loosened to increase the statistical power of the CR. The details of the TCR and the WCR are described in Section~\ref{sec:back:CRs:TCR}, while the STCR and TZCR are documented in Section~\ref{sec:back:CRs:STCR} and ~\ref{sec:back:CRs:TZCR} respectively.  Table~\ref{tab:back:SR:CRs} presents an overview of the CR selections for the TCR, WCR, and STCR corresponding to SR1, SR2, and SR3.

A likelihood fit is performed for each SR and involves the SR and the associated CRs~\cite{Baak:2014wma}. The expected number of events in each region is given by the sum over all background processes, and optionally a signal model. The normalizations of the 
$\ttbar$, $\ttbar +W/Z$,  single top, and $W$+jets backgrounds are controlled by four free parameters (normalization factors, NFs) in the fit. 
Furthermore, a signal strength parameter to normalize the cross-section of a given signal model can be included in the fit.
Each fit is based on up to five observables: the total yields in four control regions, and the total yield in one signal region.
The electron and muon channels are always added together.
To obtain a set of background predictions that are independent of the observations in the SRs, the fit can be configured 
to use only the CRs to constrain the fit parameters: the SR bins are removed from the likelihood and any potential signal contribution is neglected everywhere. This fit configuration is referred to as the background-only fit.

\begin{table}
	\centering
{\renewcommand{\arraystretch}{1.1}%
	\begin{tabular*}{\textwidth}{@{\extracolsep{\fill}}| l | ccc |}
                \hline
                \multicolumn{4}{| c |}{Common event selection} \\
                \hline
                Trigger & \multicolumn{3}{l |}{\met\ trigger} \\
                Lepton & \multicolumn{3}{l |}{exactly one signal lepton ($e$, $\mu$), no additional baseline leptons} \\
                Jets & \multicolumn{3}{l |}{at least four signal jets, and $\minDeltaPhi > 0.4$ for $i\in\{1,2\}$} \\
                Hadronic $\tau$ & \multicolumn{3}{l |}{veto events with a hadronic $\tau$ decay and $\mtTwoTau < 80$\,\GeV} \\
                \hline
		\hline
		Variable                       &  SR1  &               TCR1 /  WCR1     &               STCR1         \\ \hline
		$\ge$4 jets with $\pt >$ \,[\GeV] & $(80~50~40~40)$ &       $(80~50~40~40)$&   $(80~50~40~40)$\\
		\hline
		\met \,[\GeV]                   & $> 260$ &      {$> 200$} &    {$> 200$} \\
		\HTmissSig                 &    $>14$    &     {$>5$}  &    {$>5$} \\   
		\mt \,[\GeV]             & $>170$ & {[30,90]}    & {[30,120]} \\
		\amtTwo  \,[\GeV]          & $>175$ & {$[100,200]$} / {$>100$} & {$>200$} \\
		\topness                      &   $>6.5$     & $>6.5$ & $>6.5$ \\
		\mTopChi \,[\GeV]    & $<270$  & $<270$  & $<270$ \\
		$\DeltaR(b,\ell)$            &   $< 3.0$  &  { --} & { --} \\
		$\DeltaR(b_{1}, b_{2})$ & -- & -- & { $>1.2$} \\
		Number of $b$-tags &    $\ge1$     & $\ge1$   /{ $= 0$} & { $\ge 2$} \\
		\hline
                \hline

		                  &  SR2    &               TCR2 / WCR2     &               STCR2       \\ \hline
		$\ge$4 jets with $\pt >$ \,[\GeV]  & $(120~80~50~25)$&    $(120~80~50~25)$&      $(120~80~50~25)$\\
		\hline
		\met \,[\GeV]                   & $> 350$ &      {$> 250$} &    {$> 200$} \\
		\HTmissSig                 &    $>20$    &   {$>15$}  &    {$>5$} \\   
		\mt \,[\GeV]             & $>200$ & {[30,90]}     & {[30,120]} \\
		\amtTwo \,[\GeV]           & $>175$ & {$[100,200]$} / {$>100$} & {$>200$} \\
		$\DeltaR(b,\ell)$            &   $< 2.5$    & { --} & { --} \\
		$\DeltaR(b_{1}, b_{2})$ & -- & -- & { $>1.2$} \\
		Number of $b$-tags &    $\ge1$     & $\ge1$ / { $= 0$} & { $\ge 2$} \\
		Leading large-R jet \pt \,[\GeV]    & $>200$ &  $>200$ & $>200$ \\
		Leading large-R jet mass \,[\GeV]  & $>140$ & $>140$ & {$>0$} \\
		\dPMETsecondLargeRjet & $> 1.0$ & $> 1.0$  & $> 1.0$ \\
                \hline
                \hline

		                   &  SR3   &               TCR3 / WCR3   &               STCR3       \\ \hline
		$\ge$4 jets with $\pt >$ \,[\GeV]  & $(120~80~50~25)$ & $(120~80~50~25)$ & $(120~80~50~25)$\\
		\hline
		\met \,[\GeV]                   & $> 480$ &     {$> 280$} &    {$> 200$} \\
		\HTmissSig                 &    $>14$    &    {$>8$}  &    {$>5$} \\   
		\mt \,[\GeV]             & $>190$ & {[30,90]}   & {[30,120]} \\
		\amtTwo \,[\GeV]           & $>175$ & {$[100,200]$} / {$>100$} & {$>200$} \\
		\topness                      &  $>9.5$  &  { $>0$} &  $>9.5$ \\
		$\DeltaR(b,\ell)$            &   $< 2.8$   & { --}  & { --} \\
		$\DeltaR(b_{1}, b_{2})$ & -- & --  & { $>1.2$} \\
		Number of $b$-tags &    $\ge1$     & $\ge1$ / { $= 0$} & { $\ge 2$} \\
		Leading large-R jet \pt \,[\GeV]    & $>280$ & {$>200$} & {$>200$} \\
		Leading large-R jet mass \,[\GeV]  & $>70$ & $>70$ & $>70$ \\
                \hline
	\end{tabular*}
}
	\caption{Overview of the event selections for all SRs and the associated \ttbar\ (TCR), \Wjets\ (WCR), and $Wt$ (STCR) control regions. 
	Round brackets are used to describe lists of values and square brackets denote intervals.}
	\label{tab:back:SR:CRs}
\end{table}

\subsubsection{Top and $W$ CRs}
\label{sec:back:CRs:TCR}

The TCRs and WCRs are constructed by modifying the \mt selection in the SRs to be a window whose upper edge is near the $W$ boson mass.  
An additional upper bound on $\amtTwo$ is applied to the TCRs in order to make them orthogonal to the STCRs, described in the next section.
Furthermore, some other kinematic requirements are relaxed or removed to increase the event yields in the CRs.
The resulting selections yield 238, 102, and 121 events in TCR1, TCR2, and TCR3, respectively, which are enriched in semileptonic $\ttbar$ events with purities that vary between 75\% and 85\%.  
The WCRs are built from the TCRs by changing the $b$-jet requirement to a $b$-jet veto, and relaxing the \amtTwo\ requirement.  The $b$-jet veto suppresses $\ttbar$ events and results in a $W$+jets purity of approximately 75\% in all three regions. The selections yield 558, 135, and 352 events in WCR1, WCR2, and WCR3, respectively. 

\subsubsection{Single Top CR}
\label{sec:back:CRs:STCR}

All of the expected single-top contributions in the three SRs are in the $Wt$ channel.
This process can evade kinematic bounds from selections targeting the suppression of \ttbar.  

Nonetheless, isolating a pure sample of $Wt$ events kinematically close to the SRs is challenging due to the similarity of $Wt$ and $\ttbar$.  The $Wt$ events that pass event selections similar to those for the SRs often have a second $b$-jet within the acceptance.
The $\amtTwo$ variable is useful for discriminating between $\ttbar$ and $Wt$ because the mass of the $Wb$ system not from the resonant top quark is typically higher than for an on-shell top quark in the phase space selected by this analysis.  
Therefore, the STCRs are characterized by $\amtTwo>200$ \GeV.  
Furthermore, to increase the purity of $Wt$ and reduce the $W$+jets contamination, events are required to have two $b$-tagged jets.  
Top quark pair events often exceed the $\amtTwo$ kinematic bound when one of the two $b$-tags used in the $\amtTwo$ calculation is a jet produced from a charm quark from the $W$ decay.  
When this jet is from the same top quark as the other $b$-tagged jet, the $\Delta R$ between them tends to be smaller than for $Wt$ events that have two $b$-jets from $b$-quarks that are naturally well separated.  Therefore, to further increase the $Wt$ purity, events in the STCRs are required to have $\Delta R(b_1,b_2)>1.2$ where $b_1$ and $b_2$ are the two highest-$\pt$ $b$-tagged jets.  Figure~\ref{fig:stcr} shows distributions of the key variables for STCR1 with all requirements applied except for that on the quantity plotted. The expected purity for $Wt$ events is approximately 40\% in all three STCRs, and the selections yield 62, 71, and 45 events in STCR1, STCR2, and STCR3, respectively.

\begin{figure}
\begin{center}
\includegraphics[width=0.5\textwidth]{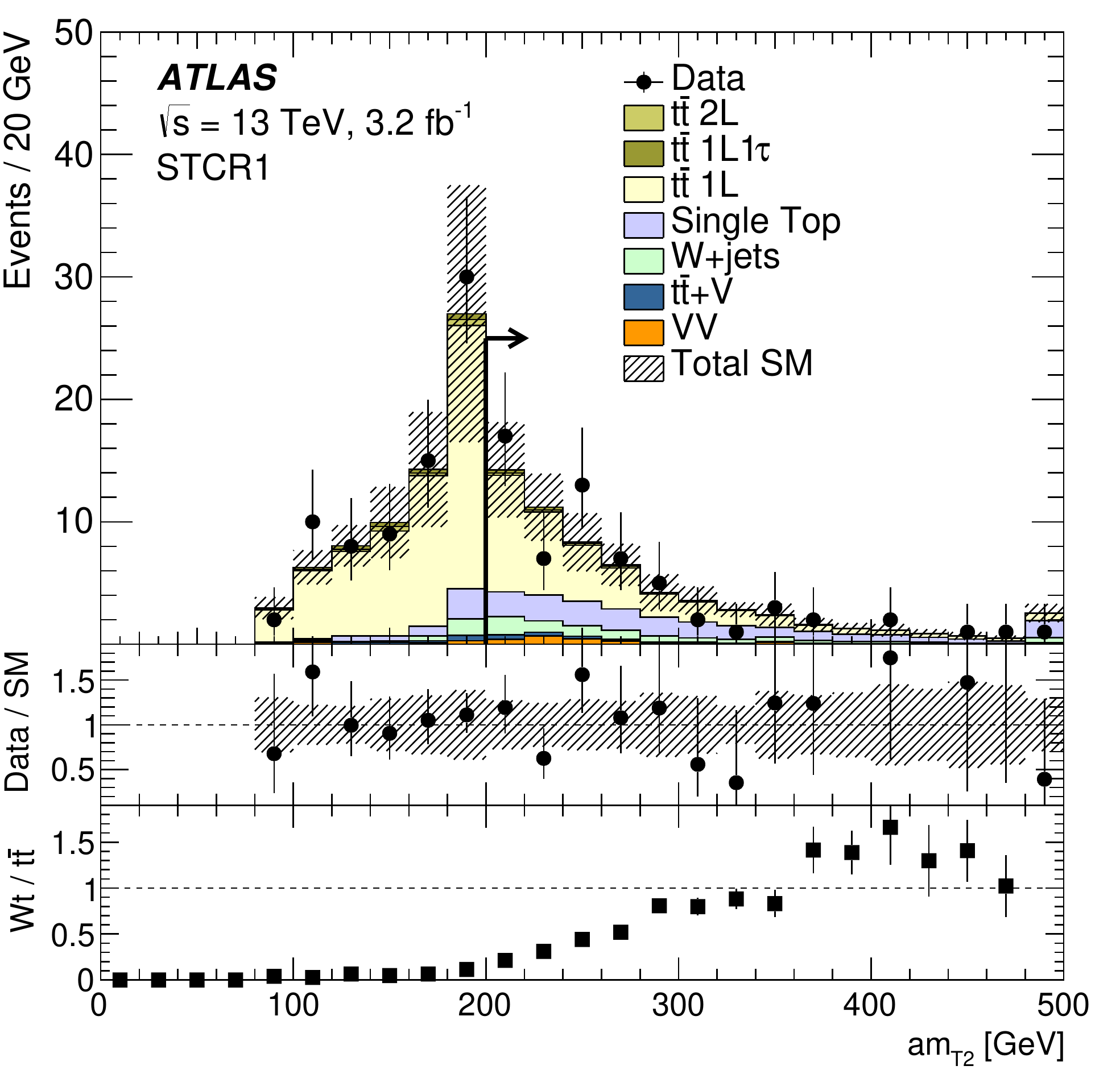}%
\includegraphics[width=0.5\textwidth]{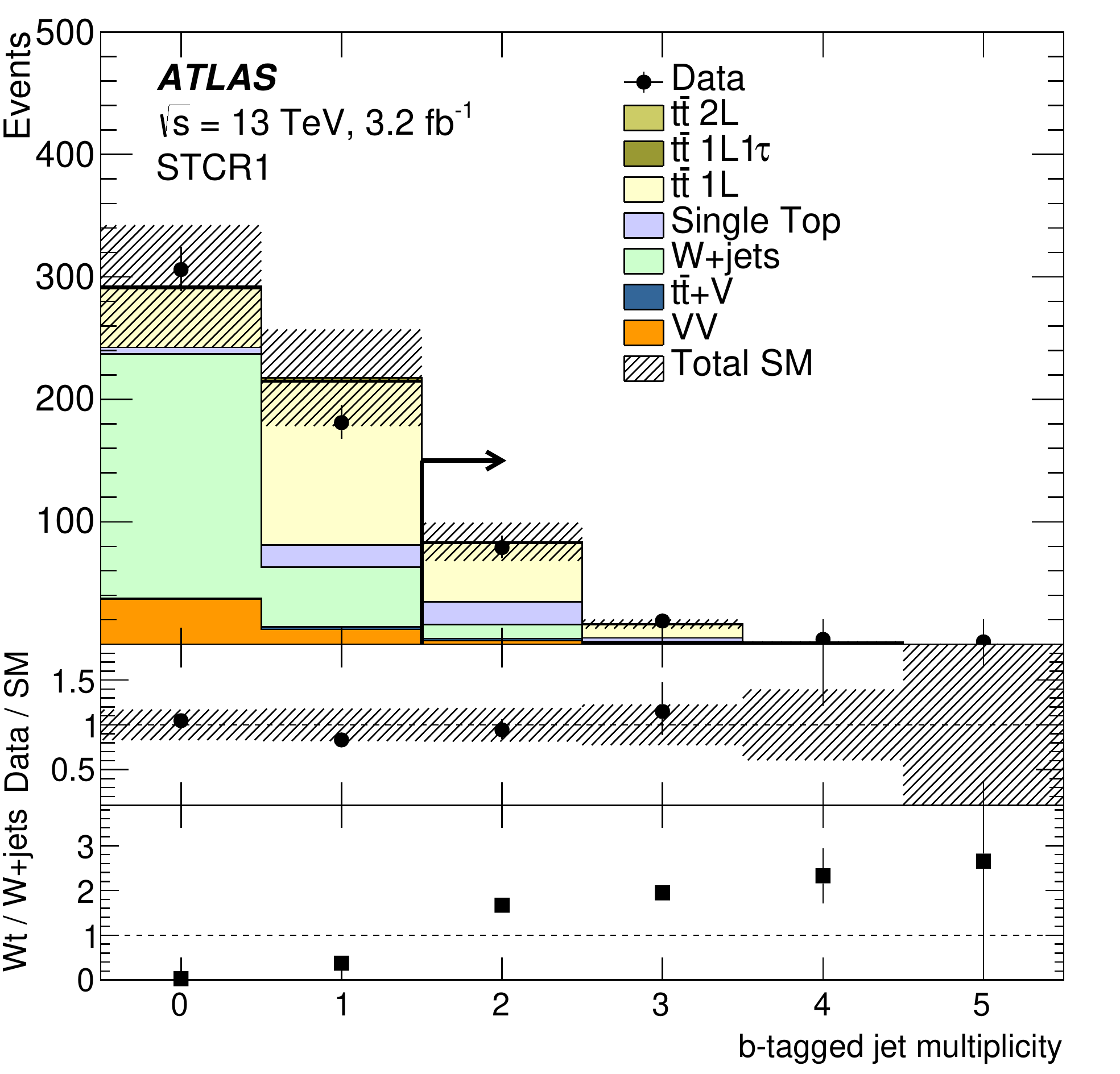}
\includegraphics[width=0.5\textwidth]{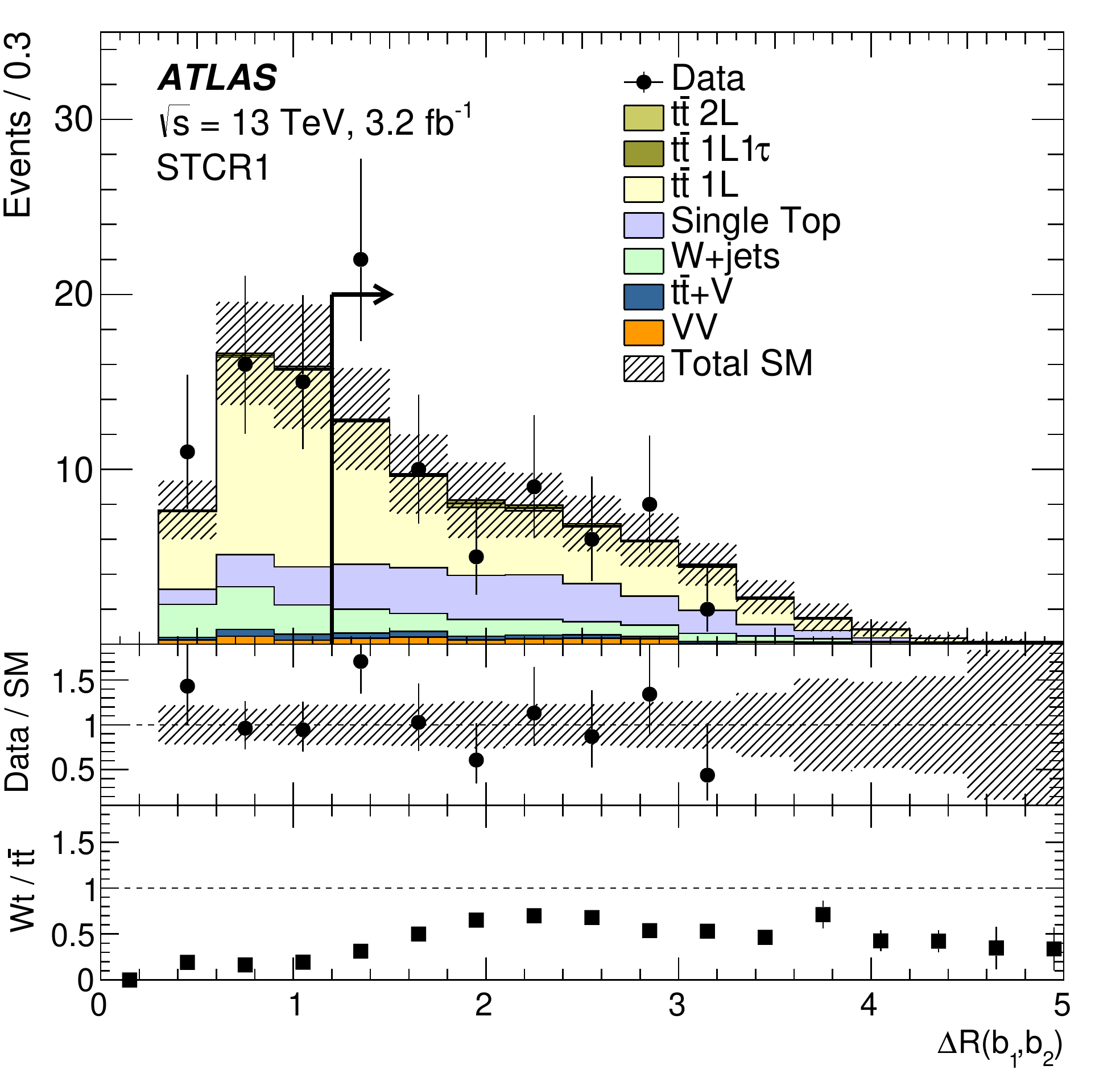}
\end{center}
\caption{Comparison of data with estimated backgrounds in the \amtTwo\ (top left), $b$-tagged jet multiplicity (top right), and 
$\Delta R(b_1,b_2)$ (bottom) distributions with the STCR1 event selection except for the requirement (indicated by an arrow) on the variable shown. 
Furthermore, the $\Delta R(b_1,b_2)$ requirement is dropped for the $b$-tagged jet multiplicity distribution.
The predicted backgrounds are scaled with the NFs documented in 
Table~\ref{tab:SR_yields}. 
The uncertainty band includes statistical and all experimental systematic uncertainties. The last bin includes overflow.
The middle panel shows the ratio of the data yield to the SM prediction, while the lower panel shows the ratio of the single-top yield to either the \ttbar\ prediction (top left and bottom) or the $W$+jets prediction (top right). The error bars in the lower panel include statistical uncertainties only.
\label{fig:stcr}}
\end{figure}

\subsubsection{$\ttbar+Z$ CR}
\label{sec:back:CRs:TZCR}

Top quark pair production in association with a $Z$ boson that decays 
into neutrinos is an irreducible background.  
The expected contributions of $\ttbar+W$ in the three SRs are less than 10\% with respect to the expected $\ttbar + Z$ yields, 
and the two processes are combined in the analysis. 
A CR using $Z$ boson decays to charged leptons is not feasible given the small branching ratio to leptons and the limited dataset available.    
However, a data-driven approach is still possible using a similar process: $\ttbar+\gamma$.  Similar techniques have been used for estimating $Z(\rightarrow\nu\bar{\nu})$+jets from $\gamma$+jets~\cite{Ask:2011xf} 
and the method was studied as a VR in the direct stop search with one lepton with Run-1 data~\cite{SUSY-2013-15}.
An event selection is constructed requiring a high-\pt photon that is then treated as \met in direct analogy to $Z\rightarrow\nu\bar{\nu}$.  

The CR is designed to minimize the differences between the two processes, in order to reduce the theoretical uncertainties in the extrapolation. The Feynman diagrams for the production of $\ttbar+Z$ and $\ttbar+\gamma$ are identical, except for a negligible production contribution where the $Z$ boson is radiated from a neutrino (the coupling is absent for photons).
The main differences arise from the $Z$ boson mass, which reduces the available phase space, causing differences in kinematic distributions. In addition, the bremsstrahlung rate for $Z$ bosons is highly suppressed at LHC energies, while there is a large contribution to the $\ttbar+\gamma$ cross-section from photons radiated from the top quark or its decay products. 
Both of these differences are mitigated if the boson \pt is larger than the $Z$ boson mass.  In this limit, the impact of the mass difference on the available phase space is reduced and the rate of photon radiation from bremsstrahlung is suppressed~\cite{Melnikov:2011ta}.  This small fraction of photons is fully accounted for in the simulation and any uncertainty in their modeling is subdominant compared to the uncertainties described in Section~\ref{sec:systematics}.  In high-\met $\ttbar+Z(\rightarrow\nu\bar{\nu})$ events, the $Z$ boson \pt is the dominant source of \met and so most $\ttbar+Z$ events in the SRs have large $Z$ boson \pt.

Two $\ttbar+\gamma$ CRs are designed to be kinematically close to SR1 and SR2/SR3.  The event selection for TZCR2 is the same as for TZCR3.  Both regions require at least one signal photon, exactly one signal lepton and no additional baseline leptons, and at least four signal jets, of which at least one must be $b$-tagged.  The two regions have the same jet \pt thresholds as the corresponding signal regions. To mimic the $Z\rightarrow\nu\bar{\nu}$ decay, the highest-\pt photon is vectorially added to $\vec{p}_\text{T}^\text{miss}$ and this sum is used to construct $\tilde{E}_\text{T}^\text{miss}=|\vec{p}_\text{T}^\text{miss}+\vec{p}_\text{T}^\gamma|$, $\tilde{m}_\text{T}$, and $\tilde{H}_\text{T,sig}^\text{miss}$.  Events entering the TZCRs are required to satisfy $\tilde{E}_\text{T}^\text{miss}>120$ \GeV, $\tilde{m}_\text{T}>100$ \GeV, and $\tilde{H}_\text{T,sig}^\text{miss}>5$ in order to bring the region kinematically closer to the SRs.  Finally, $\met < 200$ \GeV\ is imposed to ensure orthogonality between the TZCR and the other CRs and SRs.  The resulting regions have over 90\% $\ttbar+\gamma$ purity, and yield 43 and 45 events in TZCR1 and TZCR2 (=TZCR3), respectively.
Figure~\ref{fig:ttgamma} shows the distribution of $\tilde{E}_\text{T}^\text{miss}$ and $\tilde{m}_\text{T}$ in the TZCR1 corresponding to SR1 before the requirement on the plotted variable is applied.  The contribution from events not involving top quarks is negligible.  The predicted backgrounds in the figure are scaled with the NFs documented in Table~\ref{tab:SR_yields}. Without scaling, the total number of events in data is about 40\% higher than in simulation, but there is no significant evidence of mismodeling of the shapes of the various distributions within uncertainties. 

\begin{figure}
\begin{center}
\includegraphics[width=0.49\textwidth]{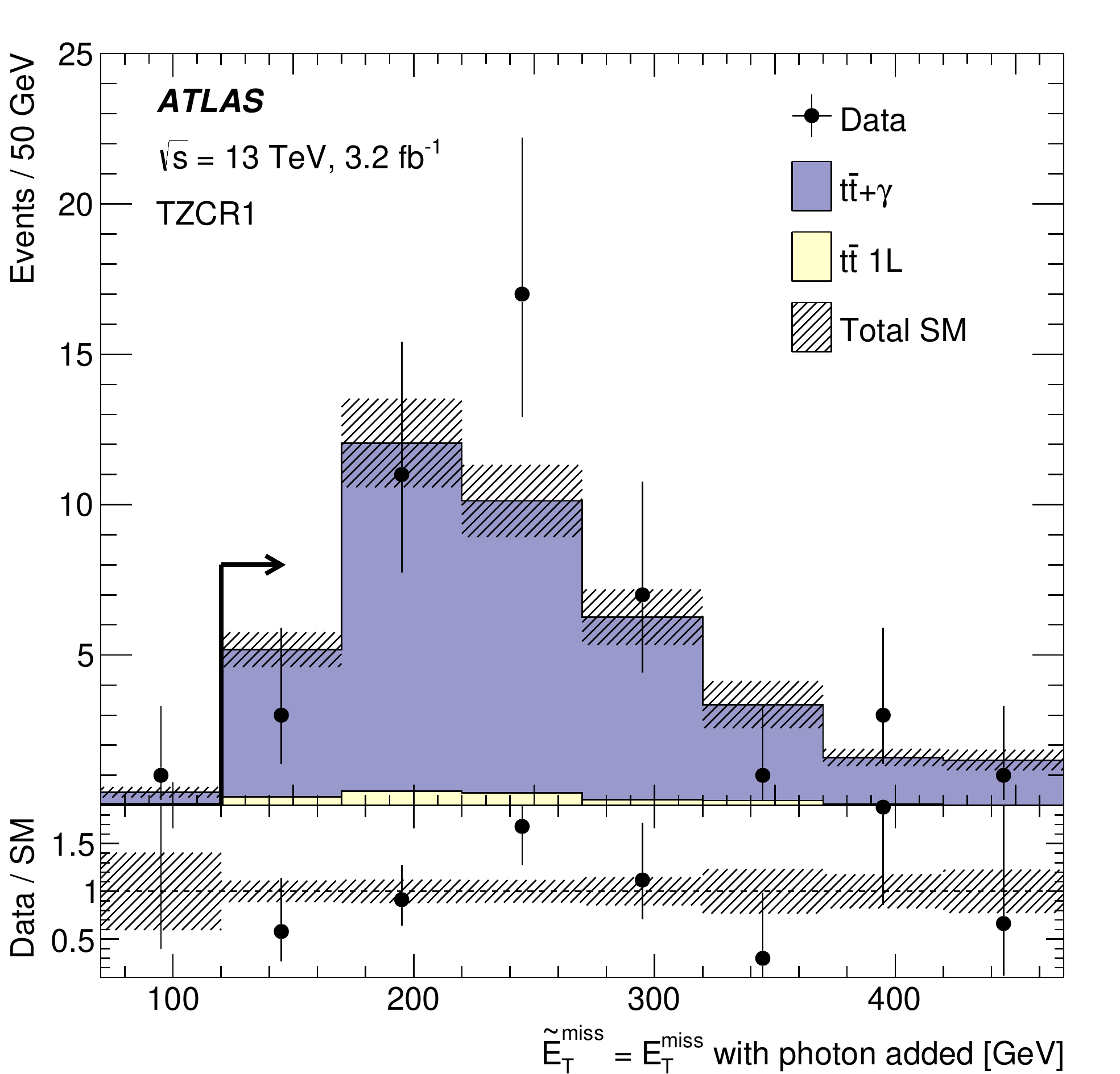}%
\includegraphics[width=0.49\textwidth]{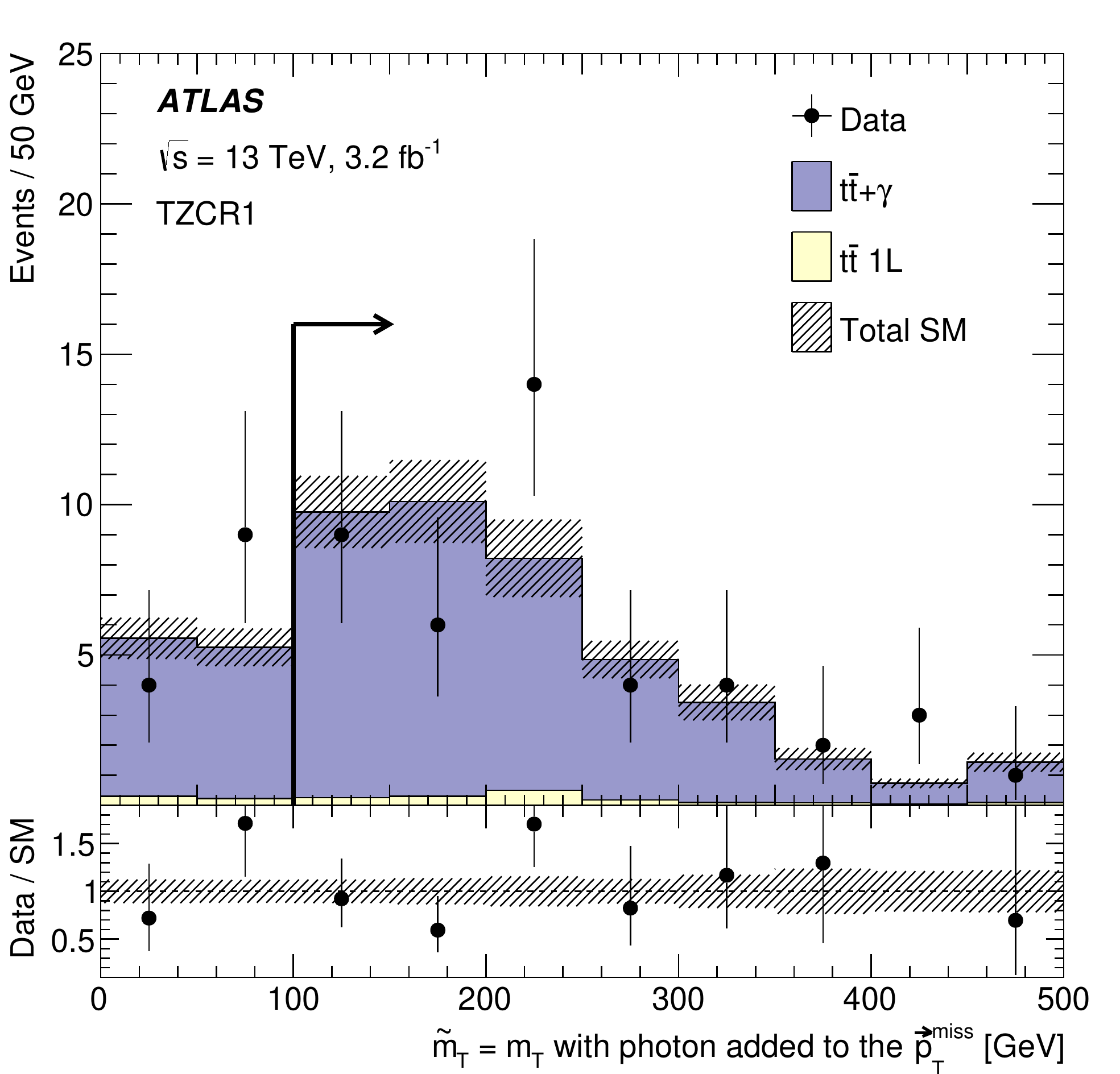}
\end{center}
\caption{Comparison of data with estimated backgrounds in the $\tilde{E}_\text{T}^\text{miss}$ and 
$\tilde{m}_\text{T}$ distributions with the TZCR1 event selection except for the requirement (indicated by an arrow) on the shown variable. The variables $\tilde{E}_\text{T}^\text{miss}$ and $\tilde{m}_\text{T}$ are constructed in the same way as \met\ and \mt but  
treating the leading photon transverse momentum as invisible. 
The predicted backgrounds are scaled with the NFs documented in 
Table~\ref{tab:SR_yields}. 
The uncertainty band includes statistical and all experimental systematic uncertainties. The last bin includes overflow.
\label{fig:ttgamma}}
\end{figure}

\subsection{Validation Regions}
\label{sec:back:VRs}

The background estimates are tested using validation regions, which 
are disjoint to both the control and signal regions.  Background normalizations determined in the control regions are extrapolated to the VRs and compared with the observed data.  Each signal region has associated validation 
regions for the \ttbar\ (TVR) and $W$+jets (WVR) processes, and these are 
constructed with the same selection as the TCR/WCR except that \mt\ is between 
$90$ and $120$\,\GeV.\footnote{A $Wt$ VR is not defined since the \mt\ range in the STCR is extended upward to $120$\,\GeV\ to accept more events.}
The validation regions are not used to constrain parameters in the fit, but 
provide a statistically independent test of the background estimates made using the CRs. In Fig.~\ref{fig:back:pulls}, 
background estimates in all the associated VRs are compared to the observed data.  The potential signal contamination in the VRs is studied for all considered signal models (and SUSY mass ranges) and found to be negligible.

\begin{figure}
\begin{center}
\includegraphics[width=0.75\textwidth]{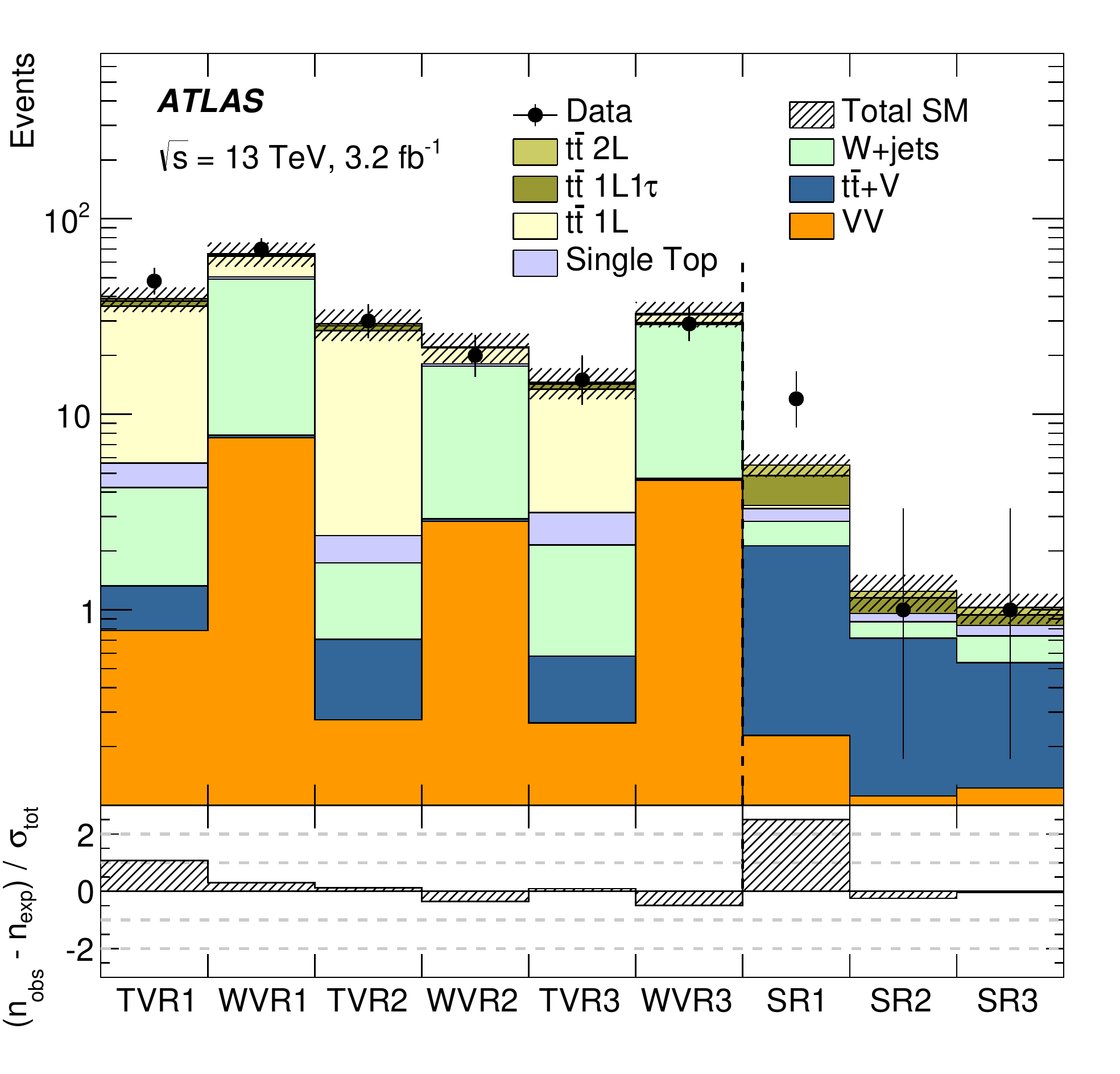}
\end{center}
\caption{Comparison of the observed data ($n_\text{obs}$) with the predicted background ($n_\text{exp}$) in the validation and 
signal regions. The background predictions are obtained using the background-only fit configuration.  
The bottom panel shows the significance of the difference between data and predicted background, where the significance is based on the total 
uncertainty ($\sigma_\text{tot}$).
\label{fig:back:pulls}} \end{figure}

A second set of validation regions, not associated with any of the three 
signal regions, is used for general monitoring purposes. Two of the more significant backgrounds are dileptonic \ttbar and lepton+hadronic $\tau$ \ttbar\ events. To pass the four-jet requirement, such events must have at least one hard jet that does not originate from the \ttbar\ decay (two hard jets for dileptonic \ttbar). The modeling of these extra jets is validated in dedicated VRs that require either two signal leptons (electron or muon) or one signal lepton and one hadronic $\tau$ candidate. In Fig.~\ref{fig:back:2L} the jet multiplicity distributions are shown for 
 event selections requiring an electron-muon pair (left) and one lepton plus one $\tau$ candidate (right).  Additional validation regions are constructed by considering (1) events with high \met, high \mt, and low \amtTwo\ for dilepton $\ttbar$ events with a lost lepton or (2) high \mt\ and a $b$-jet veto to probe the modeling of the resolution-induced $\mt$ tail in $W$+jets events (using the WVR-tail region in Fig.~\ref{fig:back:SRCRVR}).  There are no significant indications of mismodeling in any of the validation regions.

\begin{figure}
\centering
\includegraphics[width=.45\textwidth]{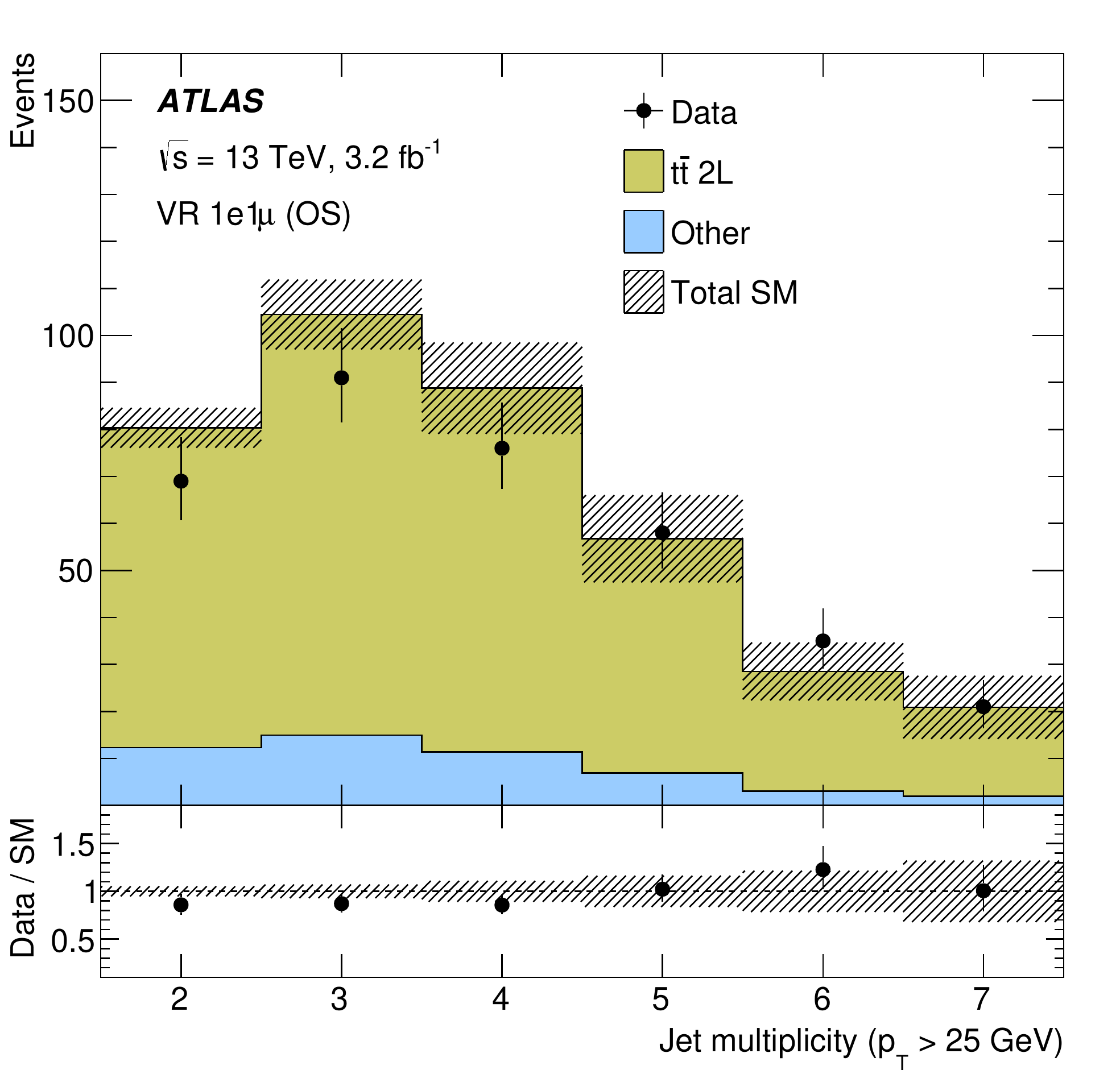}
\includegraphics[width=.45\textwidth]{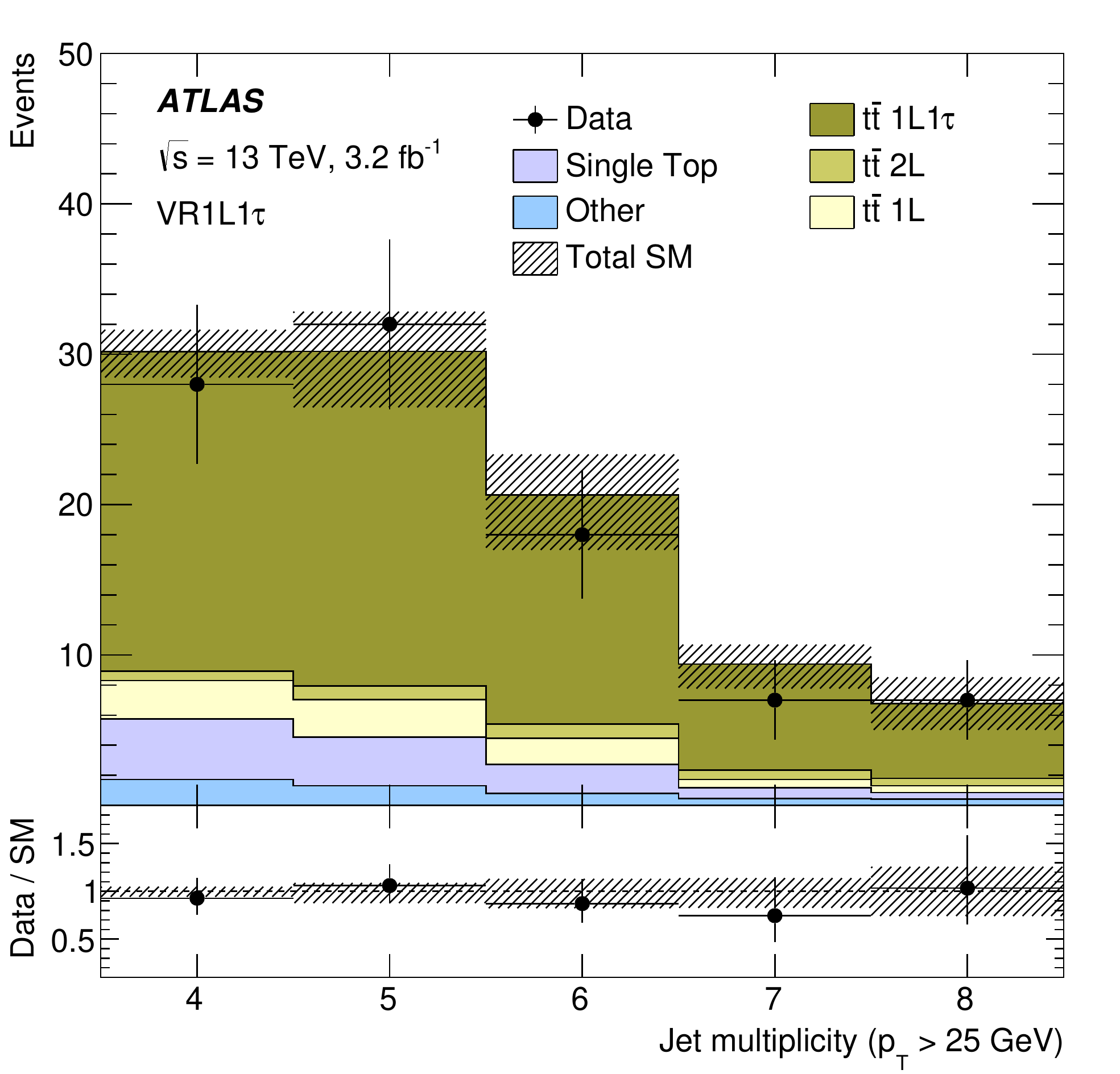}
\caption{Jet multiplicity distributions for events where exactly two signal leptons (left) or one lepton plus one $\tau$ candidate (right) are selected. No correction factors are included in the background normalizations. 
The uncertainty band includes statistical and all experimental systematic uncertainties. The last bin includes overflow.
}
\label{fig:back:2L}
\end{figure}

\section{Systematic Uncertainties}
\label{sec:systematics}

The systematic uncertainties in the signal and background estimates arise both from experimental sources and from the uncertainties in the theoretical predictions and modeling. Since the yields from the dominant background sources, \ttbar, single top, $\ttbar V$, and $W$+jets, are all obtained in dedicated control 
regions, the modeling uncertainties for these processes affect only the 
extrapolation from the CRs into the signal regions (and between the various
control regions), but not the overall normalization. 
The systematic uncertainties are included as nuisance parameters with
Gaussian constraints and profiled in the likelihood fits.

The dominant experimental uncertainties arise from imperfect knowledge of the jet 
energy scale (JES) and jet energy resolution (JER)~\cite{ATL-PHYS-PUB-2015-015},
the modeling of the $b$-tagging 
efficiencies for $b$, $c$ and light-flavor jets~\cite{ATLAS-CONF-2014-004,ATLAS-CONF-2014-046}
as well as the contribution to the \met soft-term, i.e., from tracks neither 
associated with any reconstructed objects nor identified as originating from pileup.
From these sources, the resulting uncertainties in the extrapolation factors for going from the four CRs to the SRs are  4--15\% for JES, 0--9\% for JER, 0--6\% for $b$-tagging, and 0--3\% for the \met soft-term.
Other sources of experimental uncertainty are the modeling of lepton- and photon-related quantities
(energy scales, resolutions, reconstruction and identification efficiencies, isolation, hadronic-$\tau$ identification)
and the uncertainty in the integrated luminosity. These uncertainties have a small impact on the final results.

The uncertainties in the modeling of the single-top and \ttbar backgrounds
include effects related to the MC event generator, the hadronization
and fragmentation modeling, and the amount of initial- and
final-state radiation~\cite{ATL-PHYS-PUB-2016-004}. The MC generator uncertainty is
estimated by comparing events produced with \powbox+\HERWIGpp
and with \MGaMC+\HERWIGpp.
Events generated with \powbox\ are hadronized with either
\PYTHIA\ or \HERWIGpp\ to estimate the effect from the modeling 
of the fragmentation and hadronization.
The impact of altering the amount of initial- and final-state radiation
is estimated from comparisons of \powbox+\PYTHIA samples
with different parton shower radiation, NLO radiation, and 
modified factorization and renormalization scales.  
One additional uncertainty stems from the modeling of the interference between the \ttbar\ and $Wt$ processes at NLO.
The uncertainty is estimated using inclusive $WWbb$ events, generated using \MGaMC, 
which are compared with the sum of the \ttbar\ and $Wt$ processes~\cite{ATL-PHYS-PUB-2016-004}. 
The resulting theoretical uncertainties in the extrapolation factors for going from the \ttbar\ and $Wt$ CRs to the SRs are
19--26\% for \ttbar, and 38--57\% for $Wt$ events, where the latter is dominated by the interference term.

The $\ttbar+Z$ background is normalized using the $\ttbar+\gamma$ CR and therefore there are uncertainties in both the kinematic extrapolation to the SR and in the conversion between the two processes.  As described in Section~\ref{sec:simulation}, a small correction factor is applied to the $\ttbar+\gamma$ cross-section to account for differences in the generator setup, and the same $K$-factor is used for both processes.  
A first source of uncertainty is 
estimated by coherently varying the factorization and renormalization scales between $\ttbar+Z$ and $\ttbar+\gamma$ events generated at LO
by a factor of two.  The impact of the scale choice is slightly different between $\ttbar+Z$ and $\ttbar+\gamma$, leading to a 10\% uncertainty for high-\pt bosons.  
An uncertainty due to NLO corrections is estimated by studying the kinematic dependence of the 
ratio of $\ttbar+Z$ and $\ttbar+\gamma$ $K$-factors.
This ratio is studied by computing the $K$-factor for the $\ttbar+Z$ and $\ttbar+\gamma$ processes using \MGaMC\ and OpenLoops+\SHERPA as a function of the boson \pt with a series of variations in the generator setup.  Coherently varying the factorization and renormalization scale (set to $H_\text{T}=\sum \pt$ for both LO and NLO) by a factor of two results in a  5\% uncertainty in the $K$-factor ratio.  Comparing the results obtained with the NNPDF and the CT14~\cite{Dulat:2015mca} PDF sets  changes the $K$-factor ratio by less than 2\%.  A final uncertainty of 5\% is due to the difference in $K$-factor ratios between the two generators when the same scale and PDF set is used, resulting from a different choice of electroweak scheme.  
The resulting theoretical systematic uncertainty in the extrapolation from the $\ttbar+\gamma$ CR to the SR is 12\%.

The uncertainty in the $W$+jets background from the merging of 
matrix elements and parton showers is studied by varying
the scales related to the matching scheme.
In addition, the effects of varying the renormalization, factorization, 
and resummation scales are estimated.
Since the $W$+jets background is normalized in a CR with a $b$-tagged jet veto, 
additional uncertainties in the flavor composition
of the $W$+jets events in the signal region, 
based on the uncertainties in the measurement reported 
in Ref.~\cite{STDM-2012-11} extrapolated to higher jet multiplicities, 
are applied in all regions requiring at least one $b$-tagged jet.
The resulting theoretical uncertainties in the extrapolation from the $W$+jets CR to the SR amount to about 40\%.

Since the diboson backgrounds are not normalized in a CR, the analysis is sensitive to the uncertainty in the total cross-section, estimated to be 6\%.  In addition, the estimate from the nominal \SHERPA sample is compared 
to that from a \powbox+\PYTHIA sample to account for differences 
related to the MC event  generator modeling. The resulting theoretical uncertainties for the diboson yields in the three SRs are about 50\%.

The SUSY signal cross-section uncertainty is taken from an envelope of cross-section predictions using different PDF sets and factorization and renormalization scales, as described in Ref.~\cite{Kramer:2012bx}, and the resulting uncertainties range from 13\% to 23\%.
The uncertainty in the VLQ signal cross-section is 10\%~\cite{Czakon:2011xx}.

\section{Results}
\label{sec:result}

Table~\ref{tab:SR_yields} (top part) and Fig.~\ref{fig:back:pulls} (right part) show the number of observed events together 
with the predicted number of background events in the three SRs.
The prediction is obtained using the background-only fit configuration described in Section~\ref{sec:background}.   
The SR2 and SR3 predicted yields agree well with the observed data in those regions.
Table~\ref{tab:SR_yields} (middle part) also lists the results for the four free fit parameters that control the normalization of the four main backgrounds (normalization factors, NFs), together with the associated fit uncertainties. 
To quantify the compatibility of the SM background-only hypothesis with the observations in the SRs, a profile likelihood ratio test is performed. These fits are configured to include the SR bin in the likelihood.  Table~\ref{tab:SR_yields} reports the resulting $p$-values ($p_0$), which are set to 0.5 for SR2 and SR3 since the observation lies below the prediction.  The data exceeds the background prediction in SR1 by 2.3 standard deviations. Four (eight) of the $12$ observed events are in the electron (muon) channel.
Figure~\ref{fig:n_minus_1_sr123} shows the \met and \mt distributions in SR1 for the data, for the background prediction, as well as for two representative signal models.   

The data are used to derive one-sided limits at 95\% confidence level (CL) on generic beyond-SM yields and on the considered signal models. 
The results are obtained from a profile likelihood ratio test following the
 $\text{CL}_\text{s}$ prescription~\cite{Read:2002hq}.
Model-independent upper limits on beyond-SM contributions are derived separately for each SR, where the fit is configured to include the SR and all its associated CRs.
A generic signal model is assumed that contributes only 
to the SR and for which neither experimental nor theoretical systematic uncertainties except for the luminosity uncertainty are considered.
The resulting limits, expected as well as observed, on the number of beyond-SM events are shown in the bottom rows of
Table~\ref{tab:SR_yields}. 

\begin{figure}
\begin{center}
\includegraphics[width=0.5\textwidth]{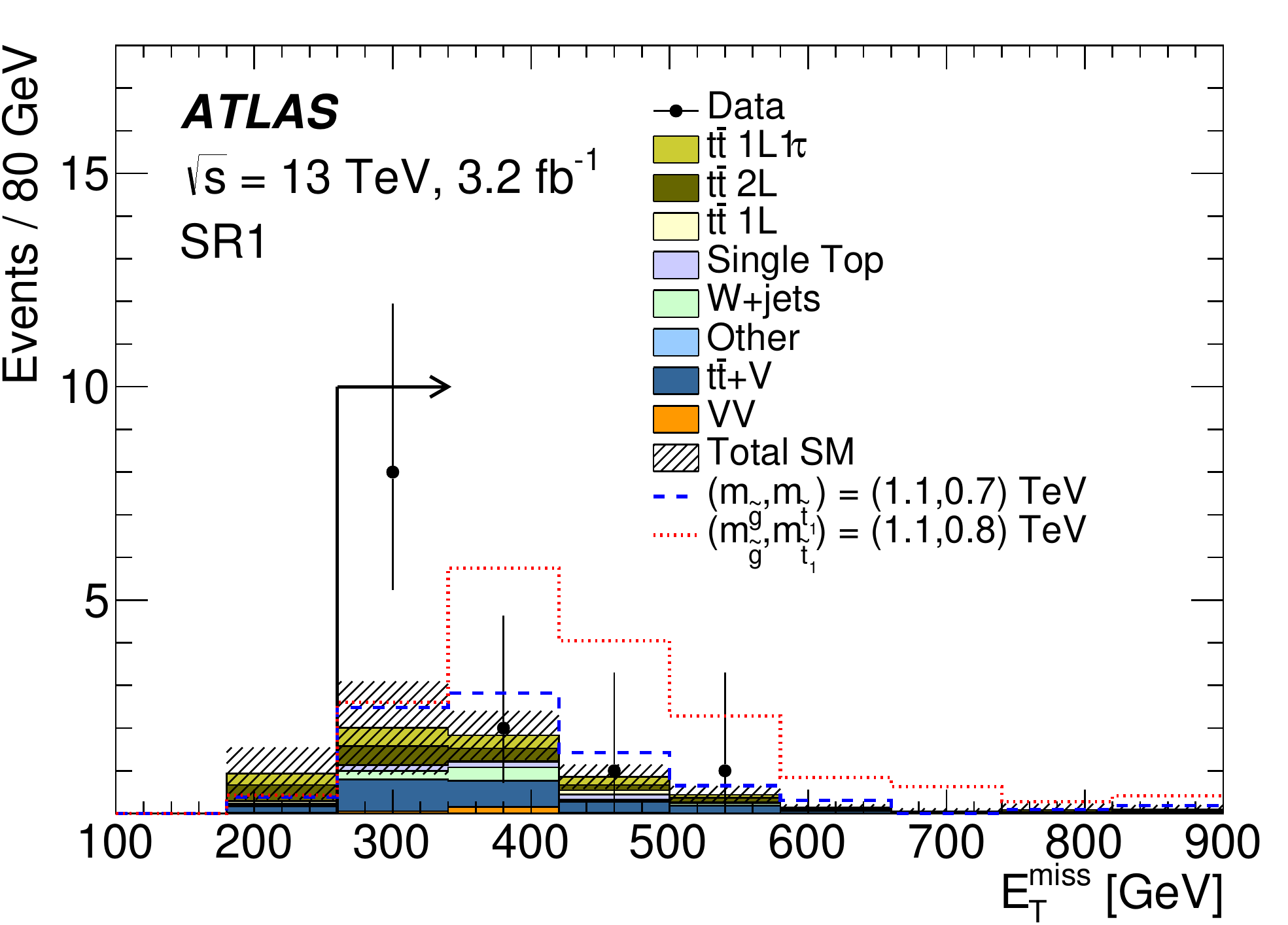}%
\includegraphics[width=0.5\textwidth]{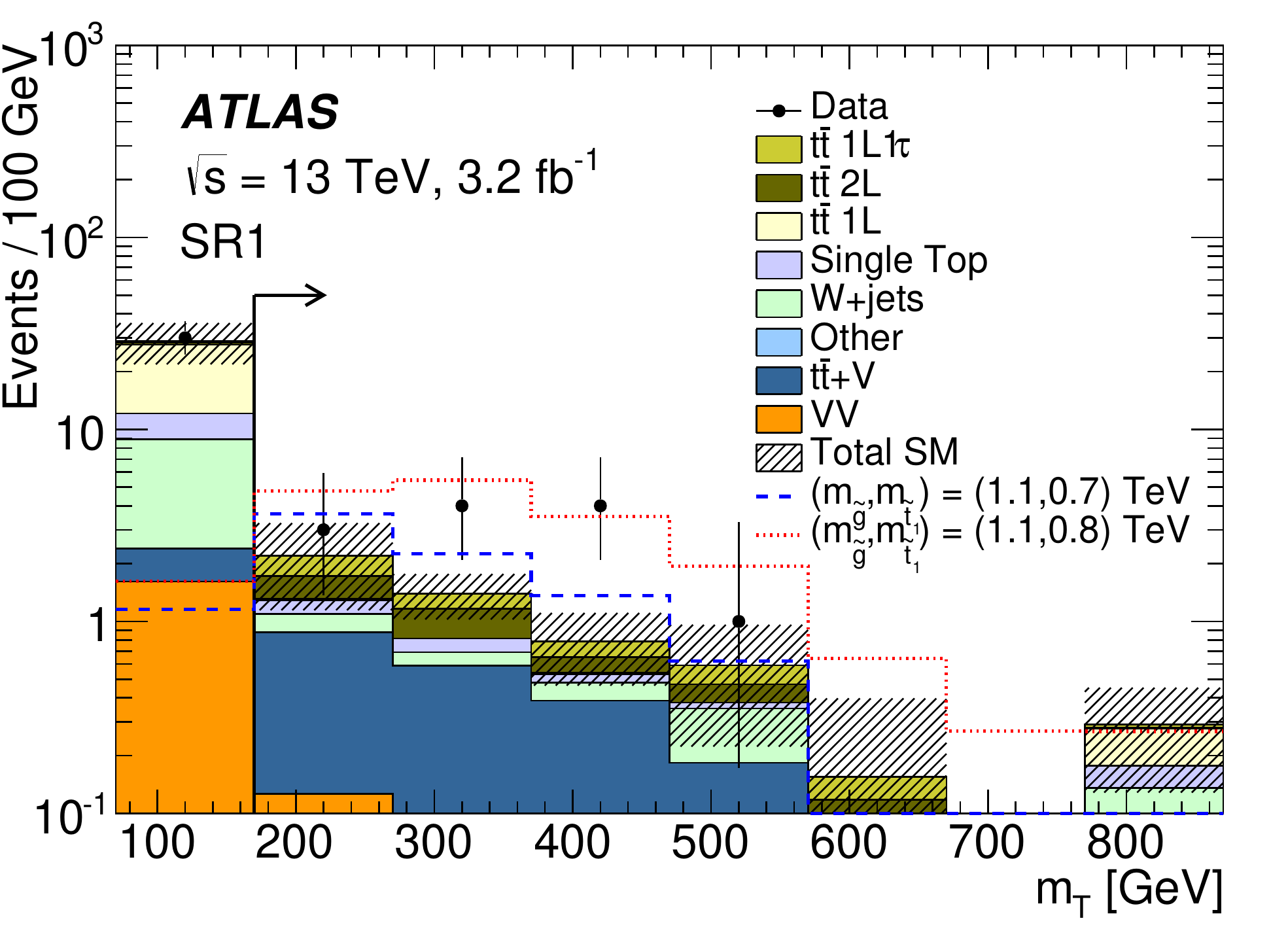}
\end{center}
\caption{The \met (left) and \mt (right) distributions in SR1. 
In each plot, the full event selection in the corresponding signal
region is applied, except for the requirement (indicated by an arrow) that is imposed 
on the variable being plotted.
The predicted backgrounds are scaled with the NFs documented in 
Table~\ref{tab:SR_yields}. 
The uncertainty band includes statistical and all experimental systematic uncertainties. The last bin contains the overflow.
Benchmark signal models are overlaid for comparison.
\label{fig:n_minus_1_sr123}}
\end{figure}

\begin{table}
\begin{center}
{\small
\setlength{\tabcolsep}{0.2pc}
{\renewcommand{\arraystretch}{1.3}%
\begin{tabular}{|l|c|c|c|}
\hline
Signal region & SR1 & SR2 & SR3 \\
\hline
Observed &  $12$               &  $1$               &  $1$               \\
\hline
Total background &  $5.50 \pm 0.72$           &  $1.25 \pm 0.26$           &  $1.03 \pm 0.18$           \\
\hline
$\ttbar$ &  $2.21 \pm 0.60$           &  $0.29 \pm 0.10$           &  $0.20 \pm 0.07$           \\
\hspace{0.5cm}(1L, 1L1$\tau$, 2L) in \% & (6, 48, 46) & (0, 58, 42) & (0, 36, 64) \\
Single top &  $0.46 \pm 0.39$           &  $0.09 \pm 0.08$           &  $0.10 \pm 0.09$           \\
$W$+jets &  $0.71 \pm 0.43$           &  $0.15_{-0.15}^{+0.19}$           &  $0.20 \pm 0.09$           \\
$\ttbar + V$ &  $1.90 \pm 0.42$           &  $0.61 \pm 0.14$           &  $0.41 \pm 0.10$           \\
Diboson &  $0.23 \pm 0.15$           &  $0.11 \pm 0.07$           &  $0.12 \pm 0.07$           \\
\hline
$\ttbar$ NF &  $ \phantom{-}1.10\pm 0.14$
 &  $ \phantom{-}1.06\pm 0.14$
 &  $ \phantom{-}0.80\pm 0.13$ \\
Single top NF &  $ \phantom{-}0.62\pm 0.46$
 &  $ \phantom{-}0.65\pm 0.49$
 &  $ \phantom{-}0.71\pm 0.42$ \\
$W$+jets NF &  $ \phantom{-}0.75\pm 0.12$
 &  $ \phantom{-}0.78\pm 0.15$
 &  $ \phantom{-}0.93\pm 0.12$ \\
$\ttbar + W/Z$ NF &  $ \phantom{-}1.42\pm 0.24$
 &  $ \phantom{-}1.45\pm 0.24$
 &  $ \phantom{-}1.46\pm 0.24$ \\
\hline
$p_0$ & 0.012 (2.3\,$\sigma$) &0.50 (0.0\,$\sigma$) &0.50 (0.0\,$\sigma$) \\
$N^{\rm limit}_{\rm non-SM}$ exp. (95\%~CL) &$6.4_{-2.0}^{+3.2}$ & $3.6_{-1.3}^{+2.3}$ & $3.5_{-1.2}^{+2.2}$ \\
$N^{\rm limit}_{\rm non-SM}$ obs. (95\%~CL) &13.3 &3.4 &3.4 \\
\hline
\end{tabular}
}
}
\end{center}
\caption{The numbers of observed events in the three SRs 
together with the expected numbers of background events and their uncertainties as predicted
by the background-only fits, the scaling factors for the background 
predictions in the fit (NF), the probabilities (represented by the 
$p_0$ values) that the observed numbers of events are compatible with 
the background-only hypothesis, as well as the 
expected and observed 95\% CL upper 
limits on the number of non-SM events.
\label{tab:SR_yields}}
\end{table}

Exclusion limits are also derived for the gluino-mediated stop and direct stop pair production models.
The signal uncertainties and potential signal contributions to all regions are taken 
into account. All uncertainties except those in the theoretical signal cross-section are 
included in the fit. Combined exclusion limits are obtained by selecting a priori 
the signal region with the lowest expected $\text{CL}_\text{s}$ value for each signal model.

Figure~\ref{fig:exclusion_contours} shows the expected and observed exclusion 
contours for both gluino-mediated and direct pair production of stops.  
The $\pm 1\,\sigma_{\rm exp}$ 
(yellow)
uncertainty band indicates the impact on the expected limit of all uncertainties included in the fit.  The $\pm 1\,\sigma_{\mathrm{th}}$ 
(dotted red) 
uncertainty lines around the observed limit illustrate the change in the
observed limit as the nominal signal cross-section is scaled up and
down by the theoretical cross-section uncertainty.  
The gap in the observed exclusion between about $600$ and $750$\,\GeV\ in the direct stop model is due to a transition between signal regions (SR1 has the best expected sensitivity up to around $750$\,\GeV\ for a massless \ninoone, beyond that SR2 has the best sensitivity) and the excess observed in SR1.
The limits are sensitive to signal model assumptions.  The gluino-mediated models have a $5$\,\GeV\ mass splitting between the stop and the neutralino and a 100\% branching ratio for $\tilde{t}\rightarrow c+\ninoone$.  The impact of varying both of these assumptions is studied for SR2 with a benchmark model characterized by masses for the gluino and the stop of $(m_{\tilde{g}},m_{\tilde{t}_1})=(1250,750)$\,\GeV.  There is a small increase in the $\text{CL}_\text{s}$ value when increasing the mass gap from $5$ to $20$\,\GeV\ and from switching between the two-body stop decay and the four-body stop decay $\tilde{t}\rightarrow bff'\ninoone$, each with 100\% branching ratio, but under all of these variations the model is excluded. The direct stop pair production limits depend on the mixing of \stopL\ and \stopR\ in forming the mass eigenstates \stopone\ and \stoptwo.  The nominal results assume that the \stopone\ is mostly the \stopR.  The stop mass limit for a massless neutralino is approximately $70$\,\GeV\ weaker when the \stopone is the \stopL.

The search for direct gluino and direct stop production can also be used to set limits in other models of physics beyond the SM that produce $\ttbar+\met$.  
Examples are third-generation leptoquarks~\cite{Dimopoulos:1979es,Dimopoulos:1979sp,Eichten:1980,Angelopoulos:1986uq,Buchmuller:1986iq,Pati:1974yy,Georgi:1974sy}, which decay into a top quark and a neutrino ($LQ\rightarrow t\nu$), and VLQ ($T$) models. 
For the former, limits on scalar $LQ\rightarrow t\nu$ are identical to limits on direct stop pair production with a massless neutralino and unpolarized top quarks.
For the latter,  simulated samples of pair-produced $T$ quarks are used to reinterpret the results.
The $T$ quark is assumed to decay in three possible ways: $T \to tZ$, $T \to tH$, and $T \to bW$. 
The search described in this paper has sensitivity mostly to the $T \to tZ$ decay mode with $Z(\rightarrow\nu\bar{\nu})$ due to the large \met\ requirements in the analysis. 
The direct $T$ pair production cross-section is higher than for stops due to additional spin states, but after accounting for the $Z(\rightarrow\nu\bar{\nu})$ branching ratio, the models have a similar predicted yield.    
For a $T$ quark with mass $800$\,\GeV\ (just beyond the Run-1 limit~\cite{EXOT-2013-17,CMS-EXO-11-005}),  a branching ratio $B\left(T \rightarrow tZ\right)$ above about 90\% is excluded.
Figure~\ref{fig:VLQ:SR3:mass} shows the exclusion limit as a function of the $T$ quark mass. Assuming a branching ratio for $T \to tZ$  of 100\%, $T$ masses up to about $850$\,\GeV\ are excluded.

\begin{figure}
\begin{center}
\includegraphics[width=0.50\textwidth]{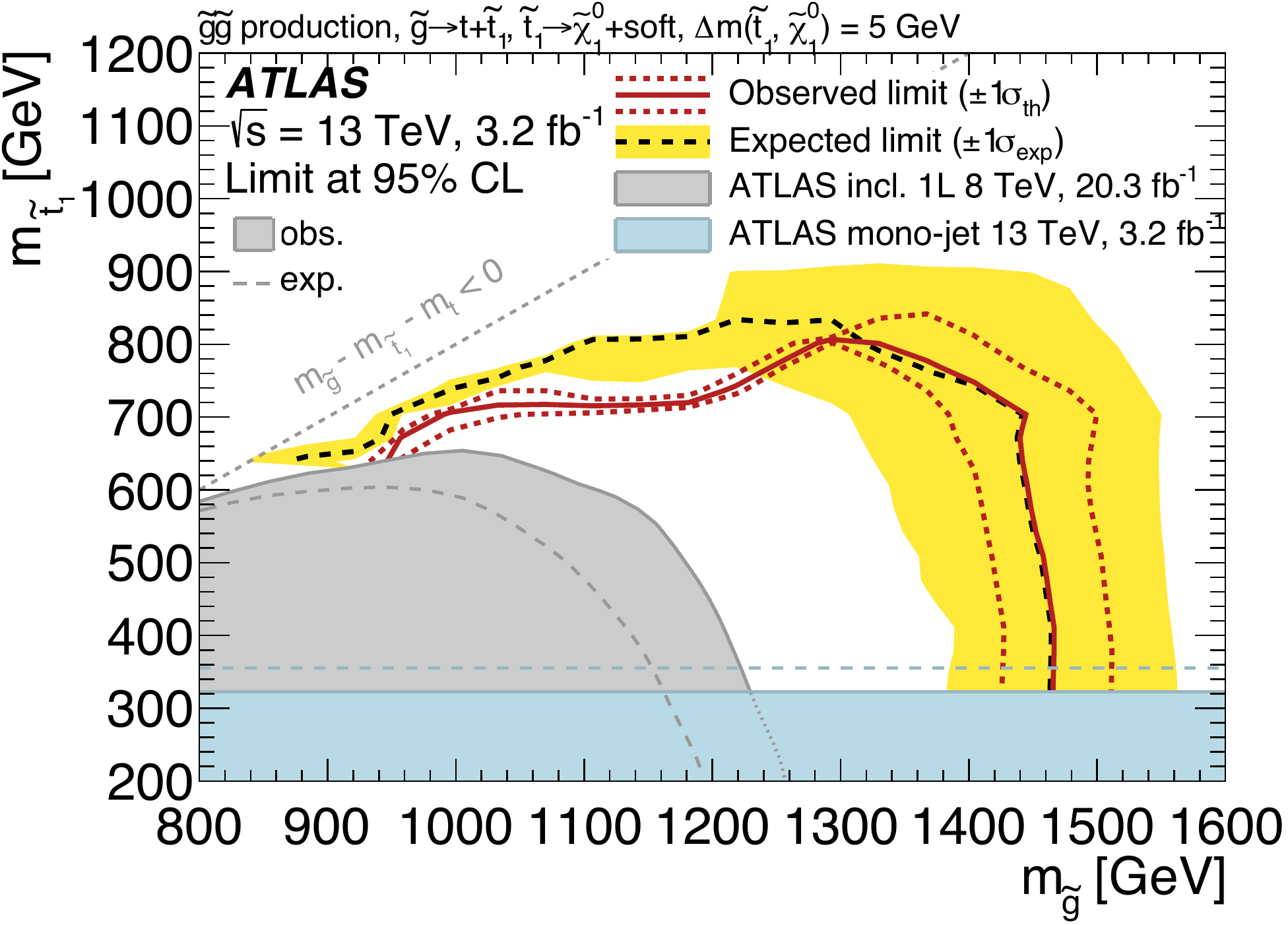}%
\includegraphics[width=0.50\textwidth]{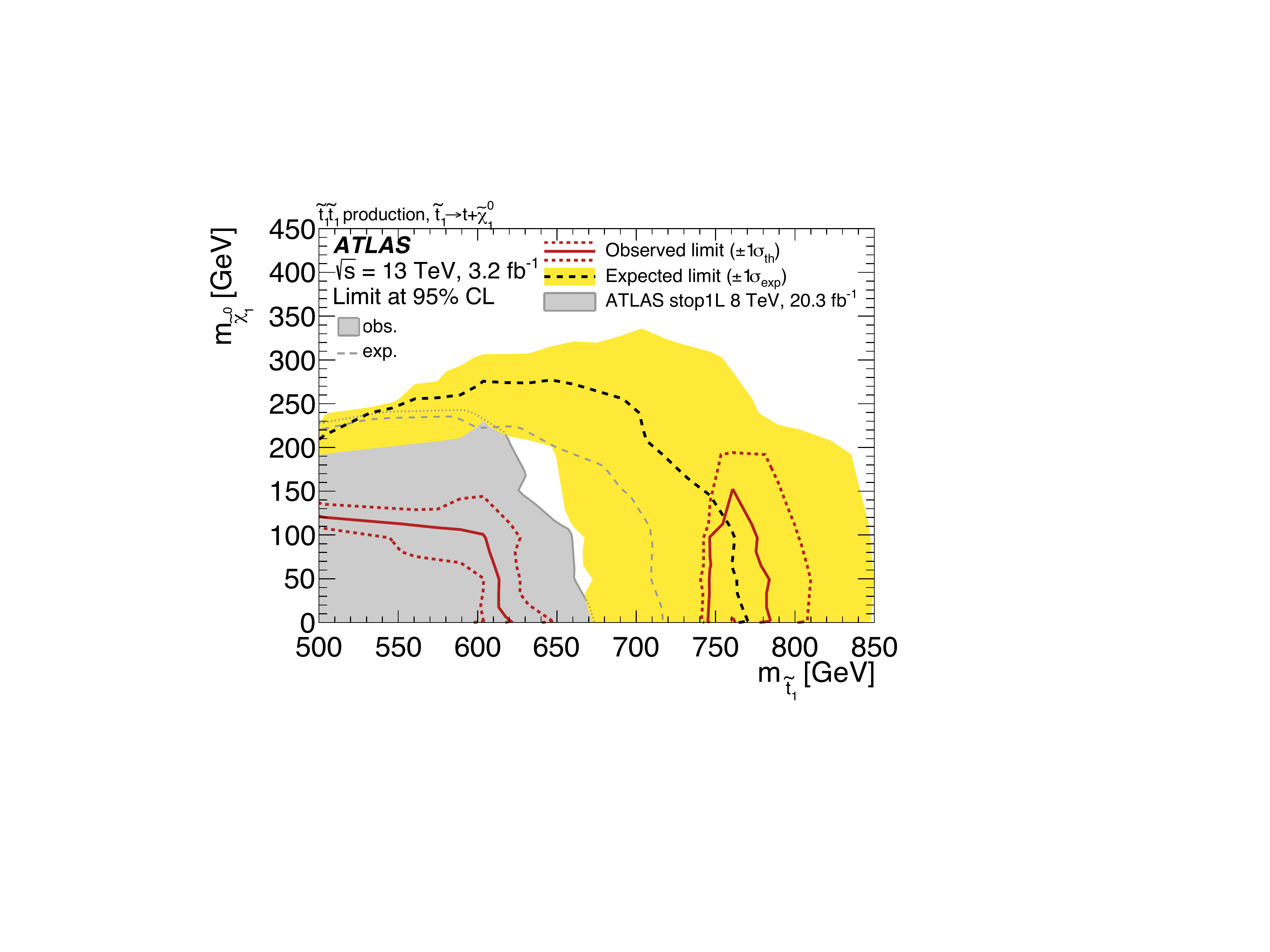}
\end{center}
\caption{Expected (black dashed) and observed (red solid) 95\% excluded regions in the plane of $m_{\tilde{g}}$ versus $m_{\stopone}$ for gluino-mediated stop production (left), and in the plane of $m_{\stopone}$ versus $m_{\ninoone}$ for direct stop pair production (right). Both scenarios assume the SUSY decays shown on the plots, each with a branching ratio of 100\%.
The gray filled areas and gray dashed lines show the observed and expected exclusion limits, respectively, from ATLAS Run-1 searches
in the inclusive one-lepton SUSY search~\cite{SUSY-2013-20} (left), and the stop search in the one-lepton channel~\cite{SUSY-2013-15} (right).
The $m_{\stopone} < 323$\,\GeV\ region for the gluino-mediated scenario (left) is excluded by the search described in Ref.~\cite{newMonojetResult}.  
The gap in the observed exclusion between about $600$ and $750$\,\GeV\ in the direct stop model is due to a transition between signal regions and the excess observed in SR1.  For any model point, the single signal region used for the observed exclusion is chosen to be the one with the best expected $\text{CL}_\text{s}$ value.
}
\label{fig:exclusion_contours}
\end{figure}

 \begin{figure}
 \begin{center}
 
 \includegraphics[width=0.50\textwidth]{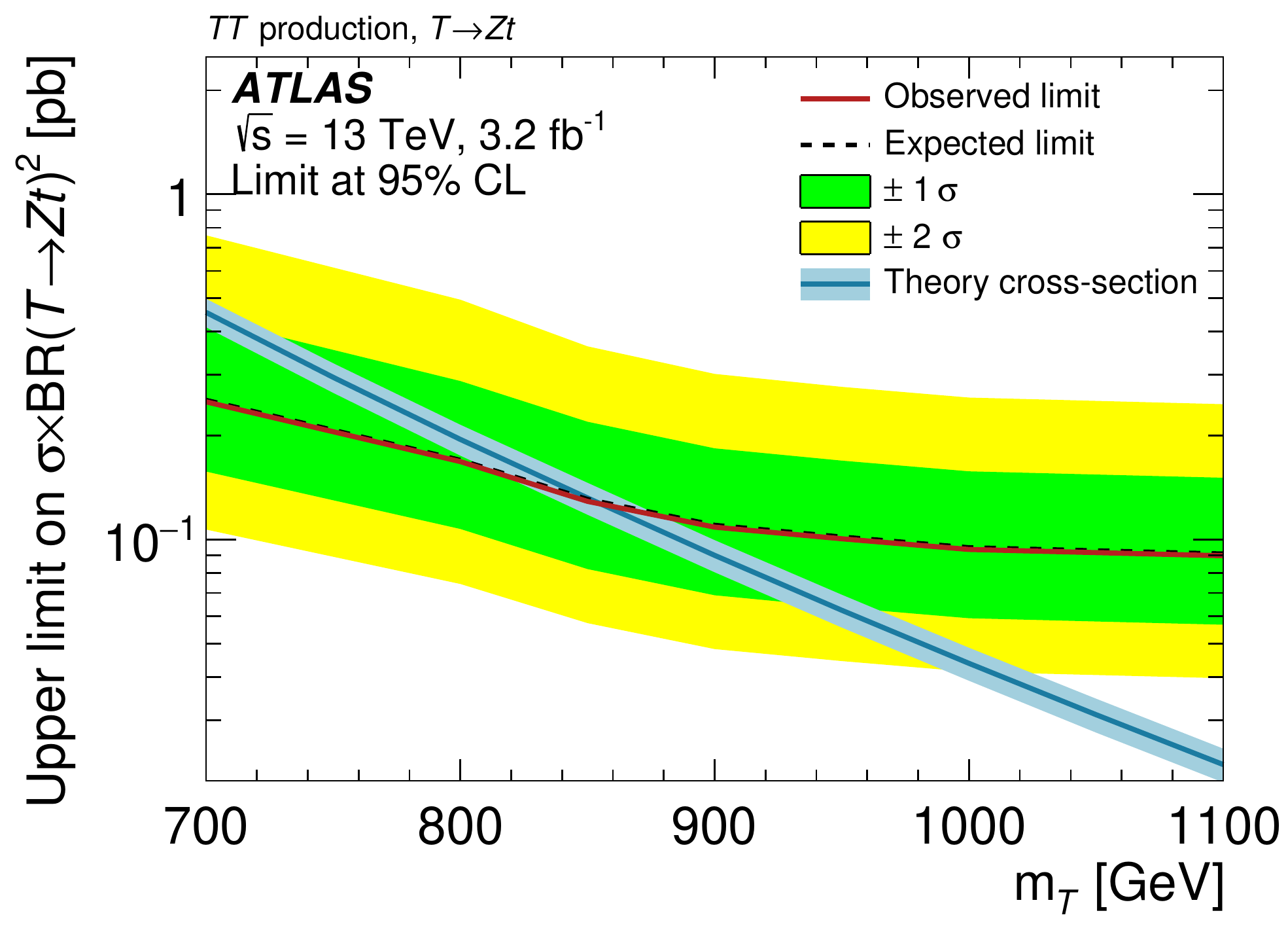} 
 \end{center}
 \caption{The observed and expected upper limits on $T$ quark pair production times the squared branching ratio for $T \to tZ$ as a function of the $T$ quark mass.  The theory cross-section is shown assuming a 100\% branching ratio for $T \to tZ$.  %
 \label{fig:VLQ:SR3:mass}}
 \end{figure}

\section{Conclusion}
\label{sec:conclusion}

This paper presents a search for pair production of gluino-mediated stops with a small mass splitting between the stop and the LSP, and direct pair production of stops decaying to two top quarks and two lightest neutralinos in final states with one isolated lepton, jets, and missing transverse momentum.  Three signal region selections are optimized for discovery in benchmark models just beyond the exclusion limits from LHC Run-1 searches with the same $\ttbar+\met$ signature. 
The search uses 3.2\,\ifb\ of LHC $pp$ collision data collected by the ATLAS experiment at a center-of-mass energy of $\sqrt{s}=13$\,\TeV.  The observed data are consistent with data-driven background estimates in all three regions.  The largest difference between data and the corresponding prediction is in the most inclusive signal region (SR1) and corresponds to 2.3 standard deviations above the estimated background.  In the absence of a significant excess, exclusion limits at 95\% CL are derived in the gluino and stop pair production models.  These extend the LHC Run-1 exclusion limits 
on the gluino mass upward to 1460\,\GeV\ in the gluino-mediated stop pair production model for low stop masses. 
For the direct stop pair production models the results expand the LHC Run-1 exclusion limits 
by excluding the stop mass region from 745 to 780\,\GeV\ for a massless lightest neutralino.
The analysis results are also reinterpreted to set exclusion limits in a model of vector-like top quarks ($T$). Assuming a branching ratio for $T \to tZ$  of 100\%, $T$ masses up to about $850$\,\GeV\ are excluded.

\section*{Acknowledgements}


We thank CERN for the very successful operation of the LHC, as well as the
support staff from our institutions without whom ATLAS could not be
operated efficiently.

We acknowledge the support of ANPCyT, Argentina; YerPhI, Armenia; ARC, Australia; BMWFW and FWF, Austria; ANAS, Azerbaijan; SSTC, Belarus; CNPq and FAPESP, Brazil; NSERC, NRC and CFI, Canada; CERN; CONICYT, Chile; CAS, MOST and NSFC, China; COLCIENCIAS, Colombia; MSMT CR, MPO CR and VSC CR, Czech Republic; DNRF and DNSRC, Denmark; IN2P3-CNRS, CEA-DSM/IRFU, France; GNSF, Georgia; BMBF, HGF, and MPG, Germany; GSRT, Greece; RGC, Hong Kong SAR, China; ISF, I-CORE and Benoziyo Center, Israel; INFN, Italy; MEXT and JSPS, Japan; CNRST, Morocco; FOM and NWO, Netherlands; RCN, Norway; MNiSW and NCN, Poland; FCT, Portugal; MNE/IFA, Romania; MES of Russia and NRC KI, Russian Federation; JINR; MESTD, Serbia; MSSR, Slovakia; ARRS and MIZ\v{S}, Slovenia; DST/NRF, South Africa; MINECO, Spain; SRC and Wallenberg Foundation, Sweden; SERI, SNSF and Cantons of Bern and Geneva, Switzerland; MOST, Taiwan; TAEK, Turkey; STFC, United Kingdom; DOE and NSF, United States of America. In addition, individual groups and members have received support from BCKDF, the Canada Council, CANARIE, CRC, Compute Canada, FQRNT, and the Ontario Innovation Trust, Canada; EPLANET, ERC, FP7, Horizon 2020 and Marie Sk{\l}odowska-Curie Actions, European Union; Investissements d'Avenir Labex and Idex, ANR, R{\'e}gion Auvergne and Fondation Partager le Savoir, France; DFG and AvH Foundation, Germany; Herakleitos, Thales and Aristeia programmes co-financed by EU-ESF and the Greek NSRF; BSF, GIF and Minerva, Israel; BRF, Norway; Generalitat de Catalunya, Generalitat Valenciana, Spain; the Royal Society and Leverhulme Trust, United Kingdom.

The crucial computing support from all WLCG partners is acknowledged
gratefully, in particular from CERN and the ATLAS Tier-1 facilities at
TRIUMF (Canada), NDGF (Denmark, Norway, Sweden), CC-IN2P3 (France),
KIT/GridKA (Germany), INFN-CNAF (Italy), NL-T1 (Netherlands), PIC (Spain),
ASGC (Taiwan), RAL (UK) and BNL (USA) and in the Tier-2 facilities
worldwide.

\printbibliography

\newpage \input{atlas_authlist}

\end{document}

%% file: atlas_authlist.tex

\begin{flushleft}
{\Large The ATLAS Collaboration}

\bigskip

M.~Aaboud$^\textrm{\scriptsize 136d}$,
G.~Aad$^\textrm{\scriptsize 87}$,
B.~Abbott$^\textrm{\scriptsize 114}$,
J.~Abdallah$^\textrm{\scriptsize 65}$,
O.~Abdinov$^\textrm{\scriptsize 12}$,
B.~Abeloos$^\textrm{\scriptsize 118}$,
R.~Aben$^\textrm{\scriptsize 108}$,
O.S.~AbouZeid$^\textrm{\scriptsize 138}$,
N.L.~Abraham$^\textrm{\scriptsize 150}$,
H.~Abramowicz$^\textrm{\scriptsize 154}$
% ...
\bigskip
\\
% ...
$^{12}$ Institute of Physics, Azerbaijan Academy of Sciences, Baku, Azerbaijan\\
% ...
$^{65}$ University of Iowa, Iowa City IA, United States of America\\
% ...
$^{87}$ CPPM, Aix-Marseille Universit{\'e} and CNRS/IN2P3, Marseille, France\\
% ...
$^{108}$ Nikhef National Institute for Subatomic Physics and University of Amsterdam, Amsterdam, Netherlands\\
% ...
$^{114}$ Homer L. Dodge Department of Physics and Astronomy, University of Oklahoma, Norman OK, United States of America\\
% ...
$^{118}$ LAL, Univ. Paris-Sud, CNRS/IN2P3, Universit{\'e} Paris-Saclay, Orsay, France\\
% ...
$^{136}$ $^{(a)}$ Facult{\'e} des Sciences Ain Chock, R{\'e}seau Universitaire de Physique des Hautes Energies - Universit{\'e} Hassan II, Casablanca; $^{(b)}$ Centre National de l'Energie des Sciences Techniques Nucleaires, Rabat; $^{(c)}$ Facult{\'e} des Sciences Semlalia, Universit{\'e} Cadi Ayyad, LPHEA-Marrakech; $^{(d)}$ Facult{\'e} des Sciences, Universit{\'e} Mohamed Premier and LPTPM, Oujda; $^{(e)}$ Facult{\'e} des sciences, Universit{\'e} Mohammed V, Rabat, Morocco\\
% ...
$^{138}$ Santa Cruz Institute for Particle Physics, University of California Santa Cruz, Santa Cruz CA, United States of America\\
% ...
$^{150}$ Department of Physics and Astronomy, University of Sussex, Brighton, United Kingdom\\
% ...
$^{154}$ Raymond and Beverly Sackler School of Physics and Astronomy, Tel Aviv University, Tel Aviv, Israel\\
% ...
$^{a}$ Also at Department of Physics, King's College London, London, United Kingdom\\
$^{b}$ Also at Institute of Physics, Azerbaijan Academy of Sciences, Baku, Azerbaijan\\
$^{c}$ Also at Novosibirsk State University, Novosibirsk, Russia\\
$^{d}$ Also at TRIUMF, Vancouver BC, Canada\\
$^{e}$ Also at Department of Physics {\&} Astronomy, University of Louisville, Louisville, KY, United States of America\\
% ...
\end{flushleft}
